\documentclass[useAMS,usenatbib]{mn2e}

\usepackage{color}
\usepackage{lscape,graphicx}
\usepackage{amsmath}
\usepackage{journal_shortcuts}
\usepackage{amssymb}
\usepackage{multirow}
\usepackage{afterpage}
\usepackage{ulem}
\usepackage{threeparttable}
\def\ga{{\rm\thinspace gauss}}

\def\h0{\hbox{{\rm H}$^0$}}
\DeclareMathAlphabet{\vib}{OML}{cmm}{m}{it}
\newcommand*{\satellite}[1]{\textit{#1}}
\newcommand*{\xmm}{\satellite{XMM-Newton}}
\newcommand*{\suzaku}{\satellite{Suzaku}}
\newcommand*{\chandra}{\satellite{Chandra}}

\newcommand{\lsim}{\mathrel{\hbox{\rlap{\lower.55ex\hbox{$\sim$}} \kern-.3em \raise.4ex \hbox{$<$}}}}
\newcommand{\gsim}{\mathrel{\hbox{\rlap{\lower.55ex\hbox{$\sim$}} \kern-.3em \raise.4ex \hbox{$>$}}}}

\title[Multiple density jumps in CIZA J2242.8+5301]{Multiple density discontinuities in the merging galaxy cluster CIZA J2242.8+5301}
\author[G.~A.~Ogrean et al.]{G.~A.~Ogrean$^{1}$\thanks{E-mail:
gogrean@hs.uni-hamburg.de}, M. Br\"uggen$^{1}$, R. van Weeren$^{2}$, H. R\"ottgering$^{3}$, A. Simionescu$^{4}$ \and M. Hoeft$^{5}$, J. H. Croston$^{6}$\\
$^{1}$Hamburger Sternwarte, University of Hamburg, Gojenbergsweg 112, 21029 Hamburg, Germany\\
$^{2}$Harvard-Smithsonian Center for Astrophysics, 60 Garden Street, Cambridge, MA 02138, USA\\
$^{3}$Leiden Observatory, Leiden University, 2300 RA Leiden, The Netherlands\\
$^{4}$Japan Aerospace Exploration Agency, 3-1-1 Yoshinodai, Sagamihara, Kanagawa 229-8510, Japan\\
$^{5}$Th\"uringer Landessternwarte, Sternwarte 5, D-07778 Tautenburg, Germany\\
$^{6}$School of Physics and Astronomy, University of Southampton, Southampton SO17 1BJ, UK}

\begin{document}

\date{Accepted xxx xxxx xx. Received xxx xxxx xx; in original form xxx xxxx xx}

\pagerange{\pageref{firstpage}--\pageref{lastpage}} \pubyear{2014}

\maketitle

\label{firstpage}

\begin{abstract}
\noindent
CIZA J2242.8+5301, a merging galaxy cluster at $z=0.19$, hosts a double-relic system and a faint radio halo. Radio observations at frequencies ranging from a few MHz to several GHz have shown that the radio spectral index at the outer edge of the N relic corresponds to a shock of Mach number $4.6\pm 1.1$, under the assumptions of diffusive shock acceleration of thermal particles in the test particle regime. Here, we present results from new \chandra\ observations of the cluster. The \chandra\ surface brightness profile across the N relic only hints to a surface brightness discontinuity ($<2\sigma$ detection). Nevertheless, our reanalysis of archival \emph{Suzaku} data indicates a temperature discontinuity across the relic that is consistent with a Mach number of $2.54_{-0.43}^{+0.64}$, in agreement with previously published results. This confirms that the Mach number at the shock traced by the N relic is much weaker than predicted from the radio. Puzzlingly, in the \emph{Chandra} data we also identify additional inner small density discontinuities both on and off the merger axis. Temperature measurements on both sides of the discontinuities do not allow us to undoubtedly determine their nature, although a shock front interpretation seems more likely. We speculate that if the inner density discontinuities are indeed shock fronts, then they are the consequence of violent relaxation of the dark matter cores of the clusters involved in the merger.
\end{abstract}


\section{Introduction}
\label{s:intro}

Galaxy clusters grow via mergers with less massive structures \citep[e.g.,][]{Kauffmann1993,Lacey1993,nfw1996}. When clusters merge, shock fronts are triggered into the intracluster medium (ICM). As shocks propagate outwards, they (re-)accelerate particles to relativistic energies. The high-energy particles interact with intracluster magnetic fields \citep[e.g.,][]{Kronberg1994} to create diffuse sources of synchrotron emission visible at radio frequencies \citep[e.g.,][and references therein]{Feretti2012}. These sources are known as radio relics, and are commonly observed at the cluster periphery \citep[e.g.,][]{Vazza2012}.

X-ray observations of clusters hosting radio relics reveal the position, extent, and strength of shock fronts. These results can be compared to radio observations to examine the correspondence between radio relics and shock fronts. Most often, relics roughly trace shock fronts present in the ICM \citep[e.g.,][]{Finoguenov2010,Macario2011,Akamatsu2013,Bourdin2013}. Sometimes, however, shock fronts are found in clusters with no radio relics \citep{Russell2010}, or shocks are significantly offset from the relics \citep{Ogrean2013d}. For a theoretical perspective on these challenges and on other open questions in our understanding of cosmic rays in galaxy clusters, see, e.g., \citet{Brunetti2014}. From an observational standpoint, to understand the connection between radio relics and shock fronts, it is essential to investigate whether the few cases that deviate from the norm are mere exceptions, or more common occurrences in merging clusters. 

Here, we present results from a 200-ks \chandra\ observation of the merging galaxy cluster CIZA J2242.8+5301. The cluster is located at $z=0.1921$, and has a $0.1-2.4$ keV luminosity of $6.8\times 10^{44}$ erg~s$^{-1}$ \citep{Kocevski2007}. \citet{vanWeeren2010} presented a radio analysis of the cluster, which revealed the presence of a double-relic system and a faint, strongly-elongated halo. Both the relics and the halo are oriented along the N-S direction, which suggests that we are observing two clusters with a N-S-oriented merger axis. Of the three main radio structures, the northern relic stands out due to its spectacular morphology: a length of 2 Mpc vs. a width of only 55 kpc. Due to this extreme shape, the relic was nicknamed the ``Sausage''. The spectral index at the outer edge of the relic predicts a Mach number of $4.6\pm 1.1$ under the assumptions of diffusive shock acceleration (DSA) of thermal particles in the linear test-particle regime. \citet{Akamatsu2013} analysed \suzaku\ observations of the cluster, and measured a temperature drop by a factor of $\sim 3$ across the ``Sausage'', which corresponds to a Mach number of $3.15\pm 0.52_{-1.20}^{+0.40}$ (the first error is the statistical error, while the following two are systematic errors). \xmm\ observations of the cluster revealed a surface brightness discontinuity east of the southern relic, and a merger geometry more complex than a binary head-on collision \citep{Ogrean2013b}. 

In the following, we further explore the surface brightness morphology and the temperature distribution within the ICM using our \chandra\ observations (PI: Ogrean) and archival \suzaku\ observations (PI: Kawahara). The paper is organised as follows: Section \ref{s:obs} presents the \chandra\ observations and the data reduction. In Section \ref{s:analysis} we analyse the X-ray morphology observed with \chandra, and in Section \ref{s:spectroscopy} we use the \suzaku\ datasets to measure the temperature distribution around the structures identified in the previous section. The results are discussed in Section \ref{s:discussion}.

We assume a flat $\Lambda$CDM universe with $H_0=70$ km\,s$^{-1}$\,Mpc$^{-1}$, $\Omega_{\rm M}=0.3$, and $\Omega_{\rm \Lambda}=0.7$. At the redshift of the cluster, 1 arcmin corresponds to 192 kpc. Throughout the paper, quoted errors are $1\sigma$ statistical errors.

\begin{figure*}
  \includegraphics[width=\textwidth]{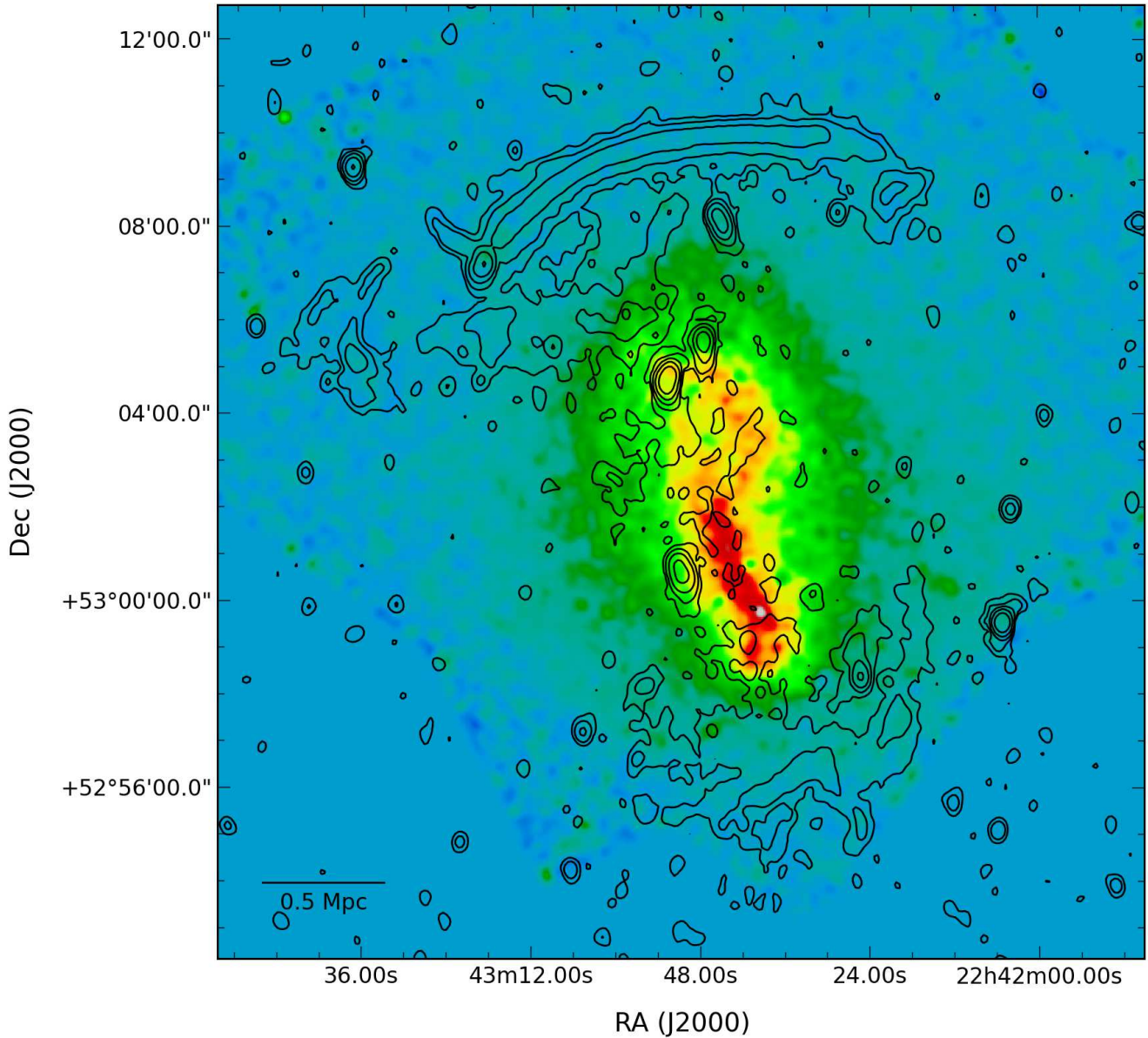}
  \caption{Surface brightness map in the energy band $0.5-7$ keV. The map was binned by a factor of 4, exposure-corrected, vignetting-corrected, instrumental background-subtracted, and smoothed with a Gaussian of kernel size 3 pixels ($\approx 6$ arcsec). Point sources were removed. Overlaid are 1.4~GHz WSRT radio contours, drawn at $[1,4,16,...]\times 70$ $\mu$Jy/beam.}
  \label{fig:sxmap-radio}
\end{figure*}

\section{Observations and data reduction}
\label{s:obs}

CIZA J2242.8+5301 was observed with the Advanced CCD Imaging Spectrometer (ACIS) on board of \chandra\ on March 29-31, 2012 (ObsID 14019) and April 21-22, 2012 (ObsID 14020), in faint mode, for a total of 200 ks (160 and 40 ks, respectively). To maximise the field of view (FOV), both observations used the four ACIS-I front-illuminated CCDs. Two of the ACIS-S CCDs -- S2 (front-illuminated) and S3 (back-illuminated) -- were also on during the observations. 

The event files of the observations were reprocessed with the newest calibration files as of June 15, 2013. The event file from ObsID 14020 was reprojected to match the coordinates of the event file from the longer observation. Time periods characterized by high background count rates are affected by soft proton flares and need to be excluded from the analysis. Therefore, we examined the $2.5-7$ keV data from the S3 chip -- the most sensitive to soft proton flares -- and used the {\sc lc\_clean} script provided by M. Markevitch to select from the light curve all time intervals during which the count rate was more than 20\% below or above the mean level; the selected time intervals were removed from the event files. 

For each observation, we created an unbinned count image and a corresponding exposure map in the energy band $0.5-7$ keV. The summed count images and summed exposure maps were used to identify point sources that contaminate the diffuse background and cluster emission. In addition, a PSF map is required in order to correctly characterise the detected sources. Therefore, for each individual observation we created a PSF map at 2.3 keV -- the same energy that was used for the exposure maps. The two PSF maps were combined with the exposure maps into a single, exposure map-weighted PSF map. Point sources were then detected with {\sc wavdetect} using wavelet radii of $2^n$ pixels, with $n=1,2,...,5$, to ensure the detection of point sources with a broad range of radii. The identified sources were confirmed visually, and detections with a significance below $3\sigma$ were discarded. For the data analysis, we used event files free of point sources.

The instrumental background was characterised using Group F stowed background event files, which are recommended for observations taken after September 9, 2009. The background event files were reprojected to match the coordinates of the cluster observations. Regions that correspond to point sources in the cluster observation were also removed from the background data. To correct for different instrumental background levels in the stowed background and observation data, we compared the $9.5-12$ keV count rates (the \chandra\ satellite's effective area drops rapidly at these high energies). The instrumental background images and spectra were renormalized to match the $9.5-12$ keV of the target observations. Between ObsID 14019 and ObsID 14020, the count rates in the high-energy band differ by less than $3\%$. 

\section{Imaging analysis}
\label{s:analysis}

To get an overall image of the cluster, we combinned $0.5-7$ keV count images and exposure maps of ObsIDs 14019 and 14020 into a single surface brightness map by dividing the summed count images to the summed exposure maps. The $0.5-7$ keV surface brightness map of the instrumental background was subtracted from the cluster surface brightness map after correcting for different instrumental background levels in the source and particle background observations, and for the different exposure times. The resulting image is shown in Figure \ref{fig:sxmap-radio}, with overlaid 1.4~GHz Westerbork Synthesis Radio Telescope (WSRT) radio contours. The details of the radio data analysis were presented by \citet{vanWeeren2010}.

\begin{figure}
  \includegraphics[width=\columnwidth]{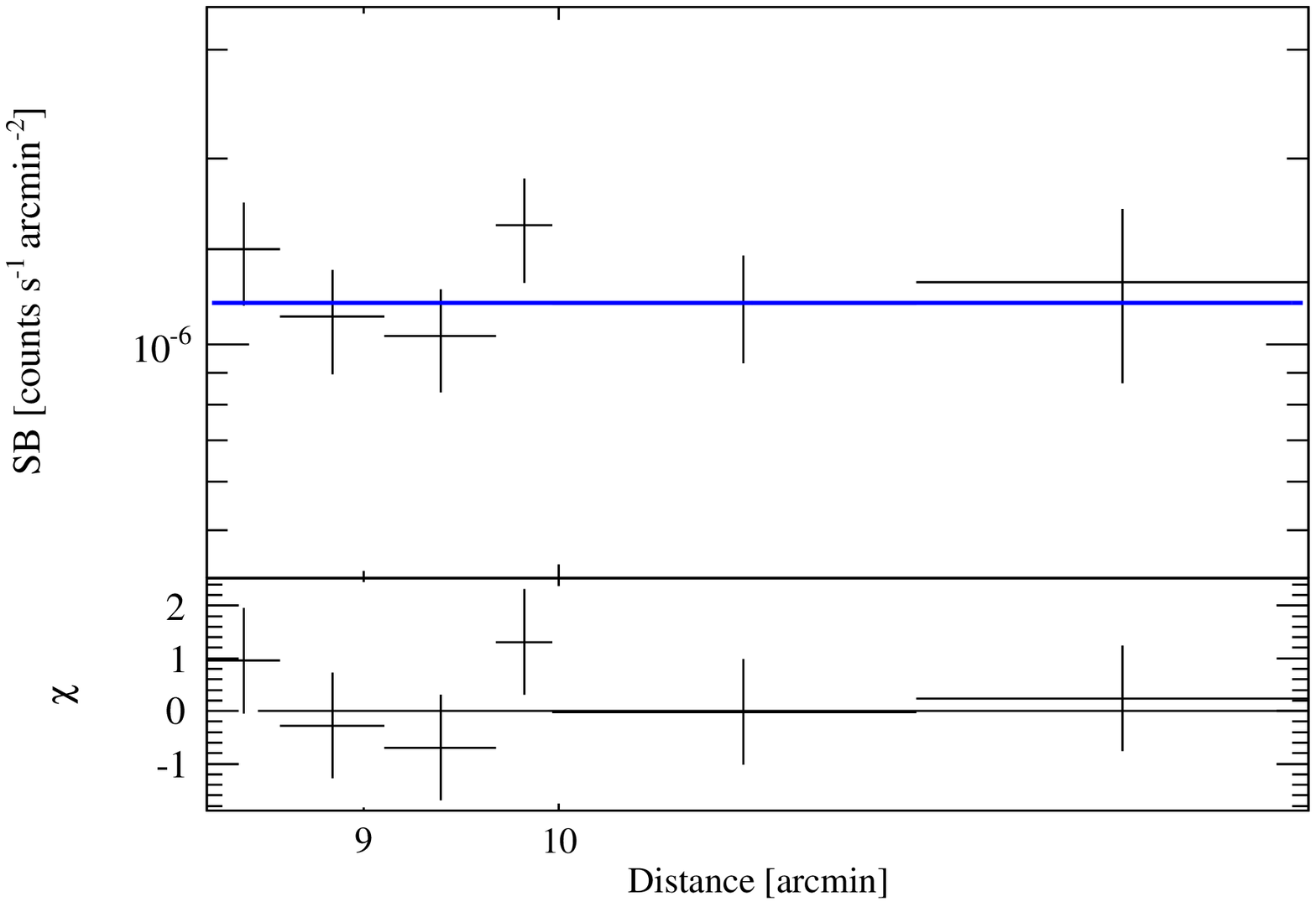}
  \caption{Background surface brightness profile, fitted with a constant (shown in blue). The profile was extracted from a partial annulus of radii 8.5 and 15 arcmin around the cluster, and binned to a target SNR of 5.}
  \label{fig:bkgsxfit}
\end{figure}

A sharp jump in surface brightness can be easily seen in the S, near the inner edge of the radio relic. Weaker discontinuities are also seen N of the merger axis and slightly to the E. A clearly linear surface brightness edge is observed to the W. Characterizing these discontinuities requires a knowledge of the sky background level. We assumed that the sky background level is constant throughout the FOV, and used \textsc{proffit} v1.1 \citep{Eckert2011b} to create a surface brightness profile in a partial annulus of radii 8.5 and 15 arcmin and opening angle 115 degrees around the cluster centre. The profile was binned to a target signal-to-noise ratio of 5, and fitted with a constant, yielding a sky background surface brightness of $(1.17\pm 0.11)\times 10^{-6}$ cts/s/arcmin$^2$. The fitted profile is shown in Figure \ref{fig:bkgsxfit}.

\begin{figure}
  \includegraphics[width=\columnwidth]{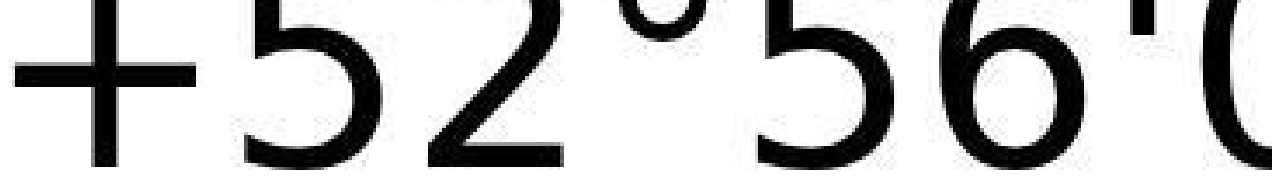}
  \caption{Sectors used for extracting surface brightness profiles.}
  \label{fig:sectors}
\end{figure}

\begin{figure}
  \includegraphics[width=\columnwidth]{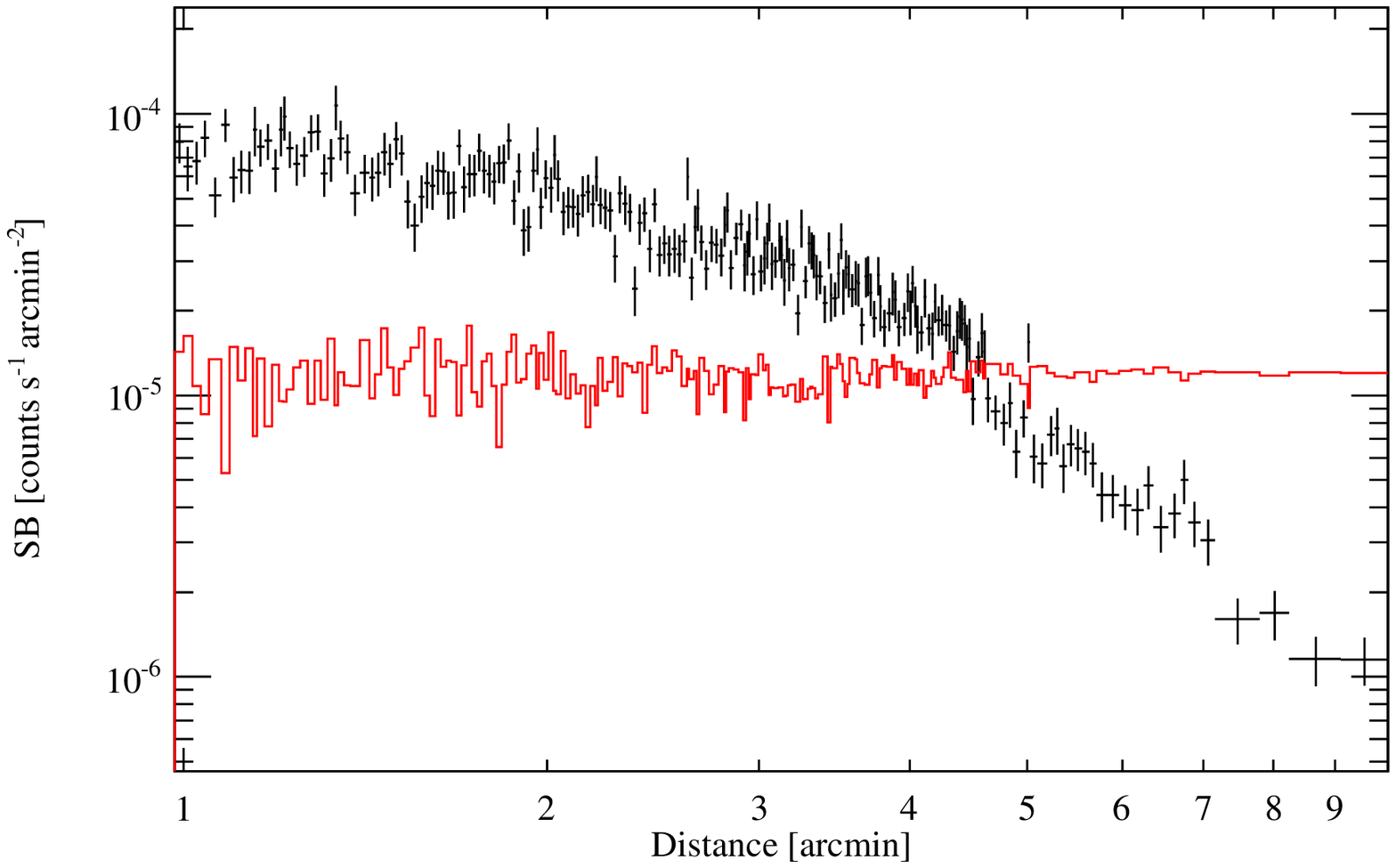}
  \includegraphics[width=\columnwidth]{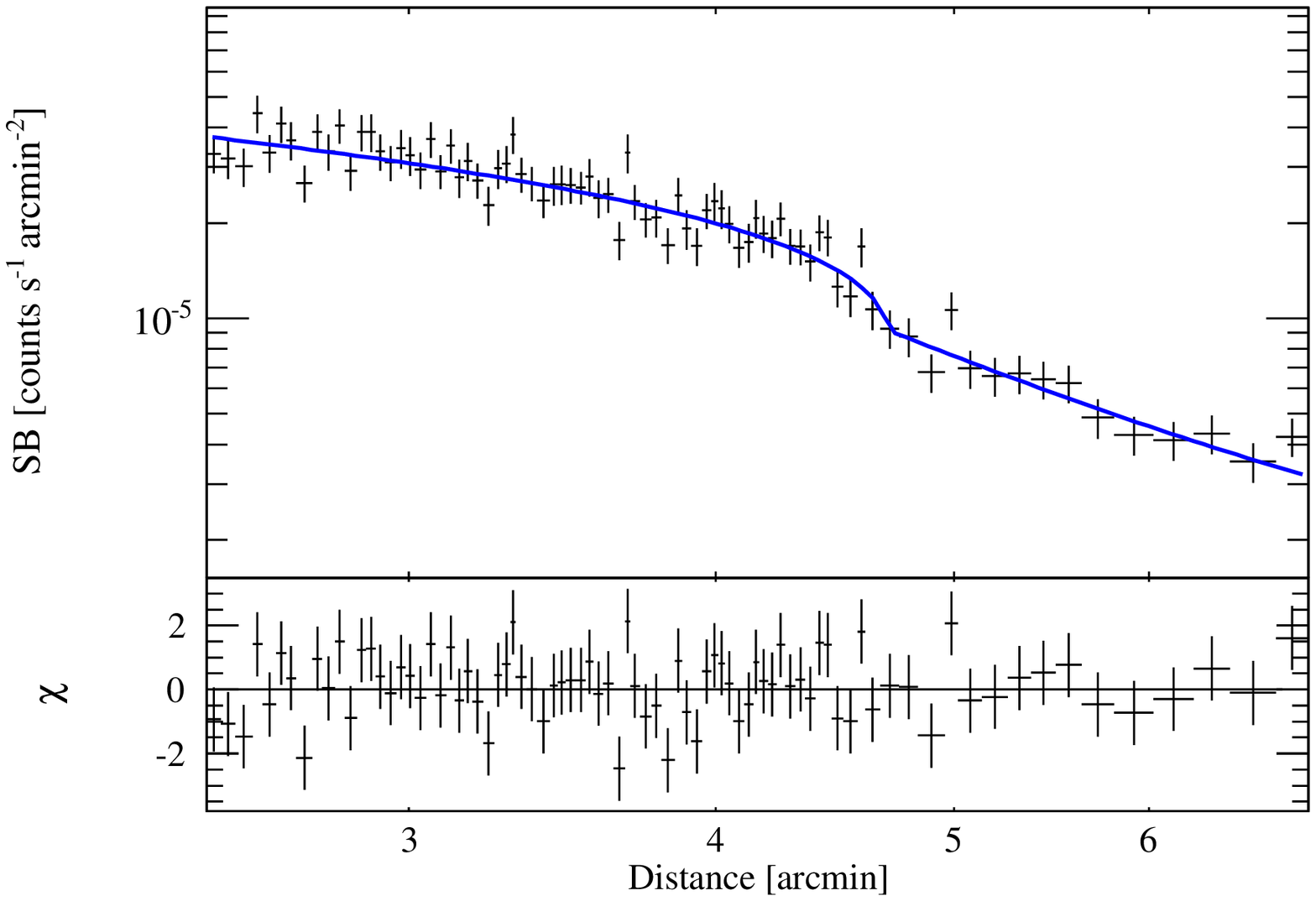}
  \caption{Surface brightness profiles extracted from the sector S1 in Figure \ref{fig:sectors}. The profiles were binned to a minimum signal-to-noise ratio of 7. On the top, we show the full surface brightness profile outside a radius of 1 arcmin, with the intrumental background (shown in red) subtracted. The bottom panel shows part of the full profile, fitted with broken power-law model (shown in blue). The discontinuity corresponds to $\mathcal{M} = 1.31_{-0.09}^{+0.12}$.}
  \label{fig:n-sx}
\end{figure}

\begin{figure}
  \includegraphics[width=\columnwidth]{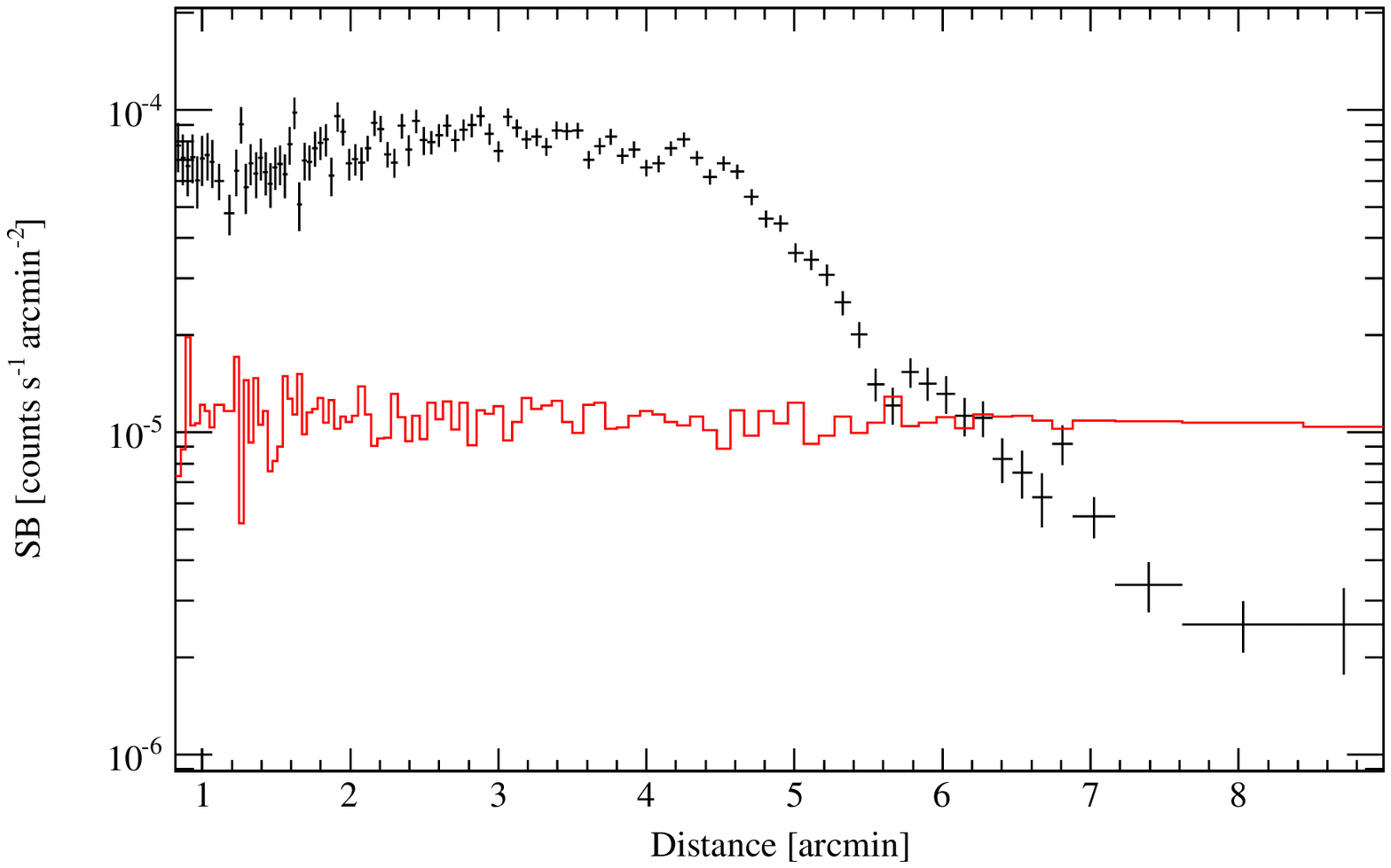}
  \includegraphics[width=\columnwidth]{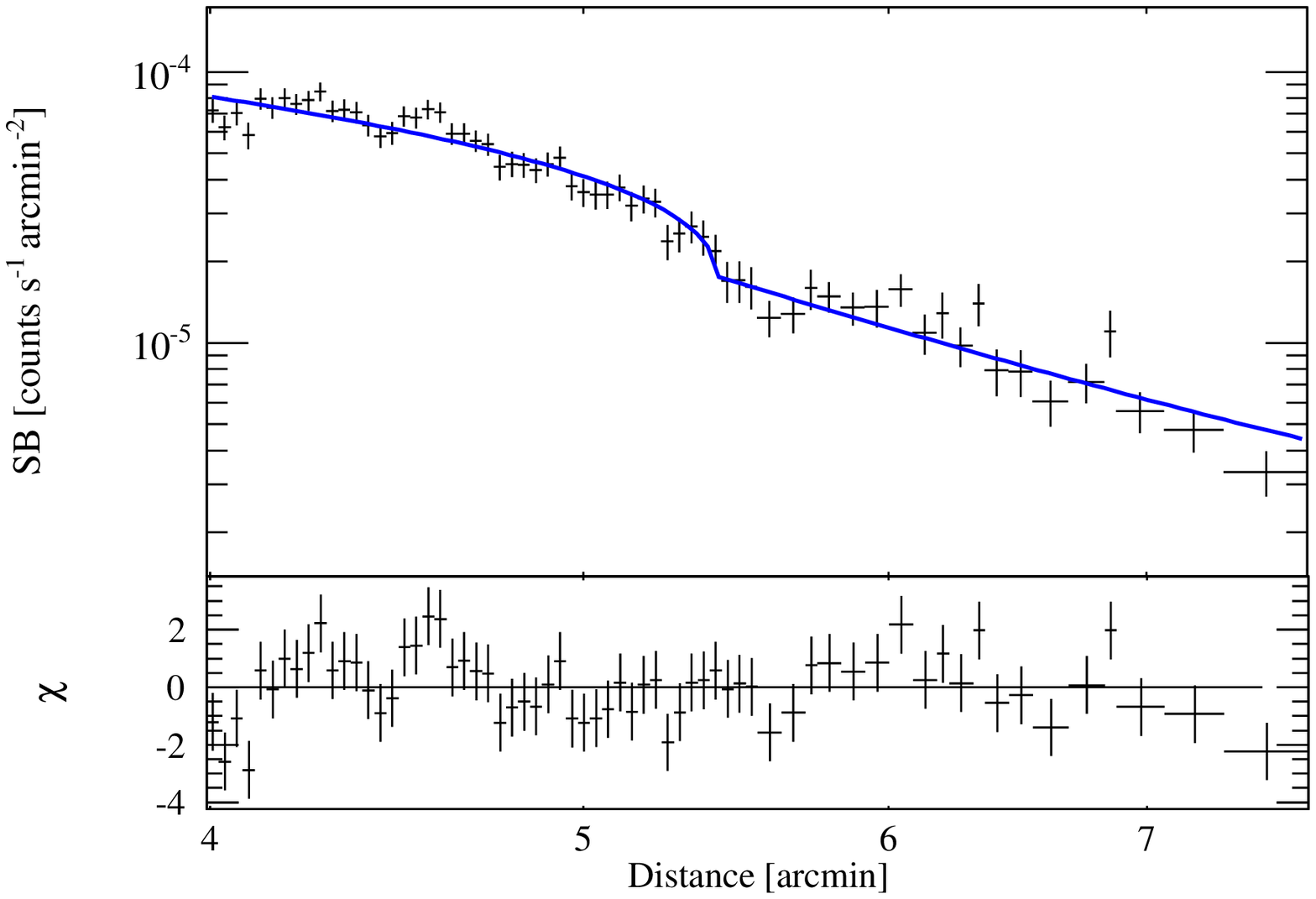}
  \caption{Surface brightness profiles extracted from the sector S2 in Figure \ref{fig:sectors}. The top panel displays the full surface brightness profile outside a radius of 1 arcmin, with the instrumental background (shown in red) subtracted. A clear discontinuity in the shape of the profile is seen near $5.5$ arcmin. At the bottom, we show the profile around this discontinuity, fitted with a broken power-law model (shown in blue). The discontinuity corresponds to $\mathcal{M} = 1.43_{-0.08}^{+0.09}$.}
  \label{fig:s-sx}
\end{figure}

Further, we extracted surface brightness profiles across the discontinuities seen in Figure \ref{fig:sxmap-radio}. The sectors are shown in Figure \ref{fig:sectors}.

Figure \ref{fig:n-sx} presents the northern surface brightness profile, binned to a target SNR of 7. Two discontinuities can be seen near 4.7 and 7 arcmin. We attempted to fit each of them with a surface brightness model consistent with an underlying broken power-law density profile:
\begin{eqnarray}
  n_1 & = & C\, n_0\, \left(\frac{r}{r_{\rm sh}}\right)^{-\alpha_1} \,\,\, , \,\,\,\,\, r \le r_{\rm sh} \nonumber \\
  n_2 & = & n_0\, \left(\frac{r}{r_{\rm sh}}\right)^{-\alpha_2} \,\,\, , \,\,\,\,\, r > r_{\rm sh} \nonumber
\end{eqnarray}
where $n$ is the electron number density, $C$ is the density compression factor, $n_0$ is the density immediately ahead of the putative outward-moving shock front, $\alpha_1$ and $\alpha_2$ are the power-law indices, $r$ is the radius from the cluster centre, and $r_{\rm sh}$ is the radius corresponding to the putative shock front. Indices $1$ and $2$ are associated with the post-shock and pre-shock regions, respectively. To derive the best-fit to the surface brightness profile, the broken power-law density profile was projected along the line of sight, under the assumption of spherical symmetry.

Unfortunately, the net count statistics beyond the outer N discontinuity are very poor, and we were only able to detect a shock at slightly more than $1\sigma$ confidence level. Therefore, the fit to the outer part of the profile is not presented here. To determine the jump strength at the N inner discontinuity, we fitted the inner part of the profile in the distance range $2.5-6.5$ arcmin. The sky background level was fixed to $(1.17\pm 0.11)\times 10^{-6}$ cts/s/arcmin$^2$, as determined above. All the other parameters were left free in the fit. The results of the fit are shown in Table \ref{tab:sxfits}. The density compression factor corresponds to a Mach number of $1.31_{-0.09}^{+0.12}$. 

The surface brightness profile along the S sector is shown in Figure \ref{fig:s-sx}. A discontinuity is easily seen near $5.5$ arcmin. Similarly to the N profile, this discontinuity was fitted with a broken power-law, leaving all the parameters free in the fit. The best-fit model is summarized in Table \ref{tab:sxfits}. The density discontinuity corresponds to a Mach number of $1.43_{-0.08}^{+0.09}$.

\begin{figure}
  \includegraphics[width=\columnwidth]{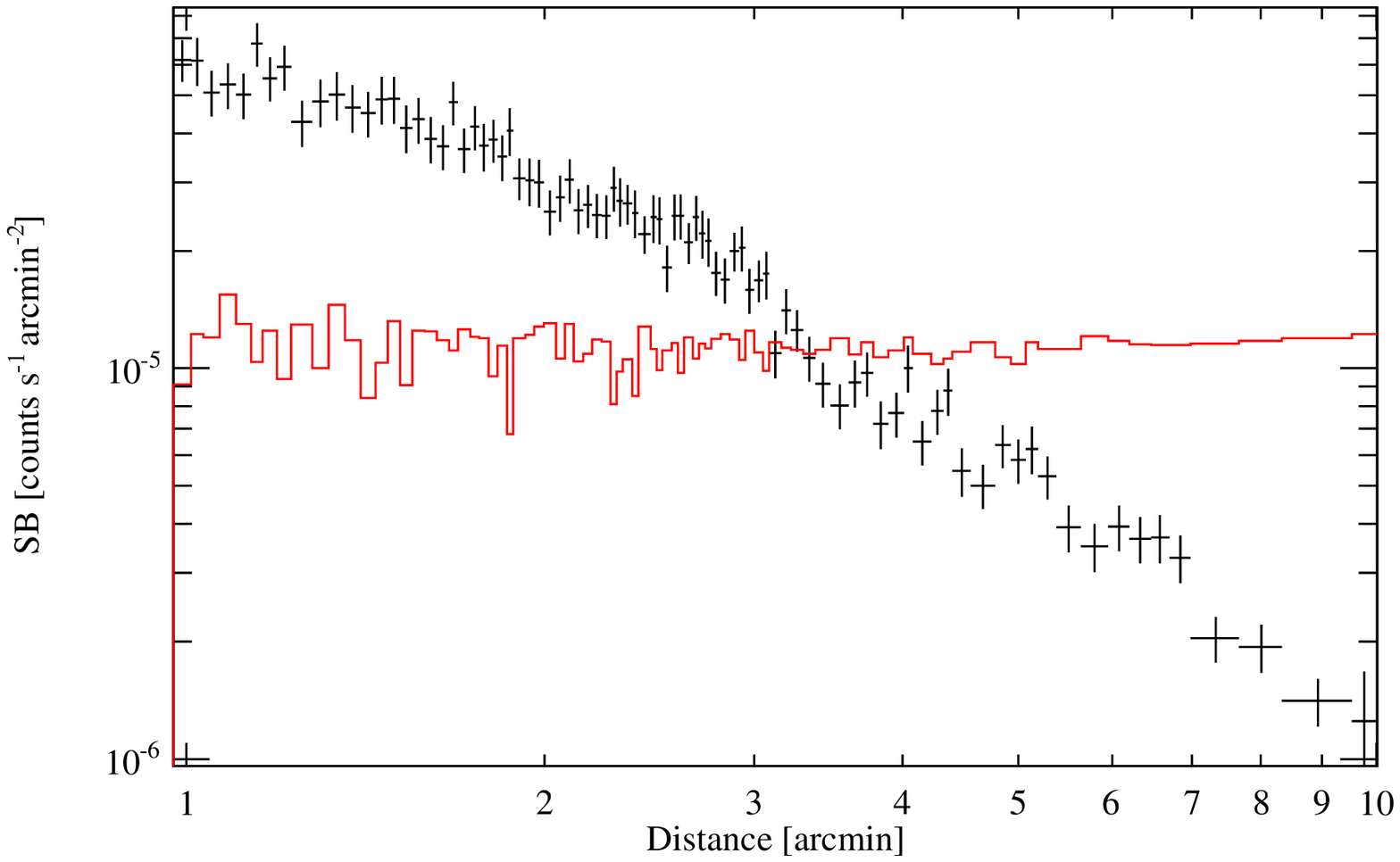}
  \includegraphics[width=\columnwidth]{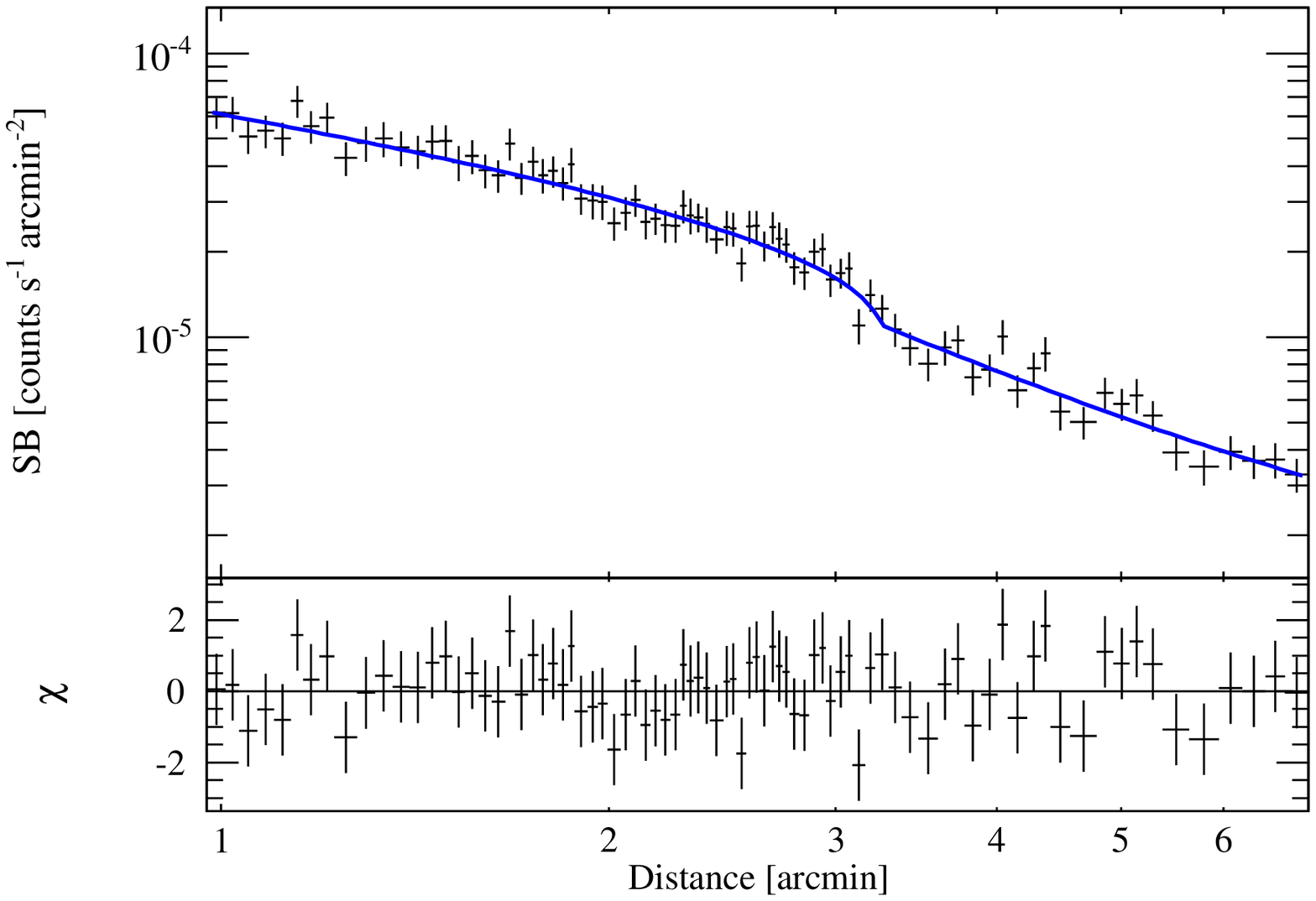}
  \caption{Surface brightness profiles extracted from sector S3 in Figure \ref{fig:sectors}. The profiles were binned to a minimum signal-to-noise ratio of 7. On the top, we show the full surface brightness profile outside a radius of 1 arcmin, with the intrumental background (shown in red) subtracted. The bottom panel presents part of the profile, fitted with a broken power-law model (shown in blue). The discontinuity corresponds to $\mathcal{M}=1.25_{-0.06}^{+0.07}$.}
  \label{fig:w-sx}
\end{figure}

\begin{figure}
  \includegraphics[width=\columnwidth]{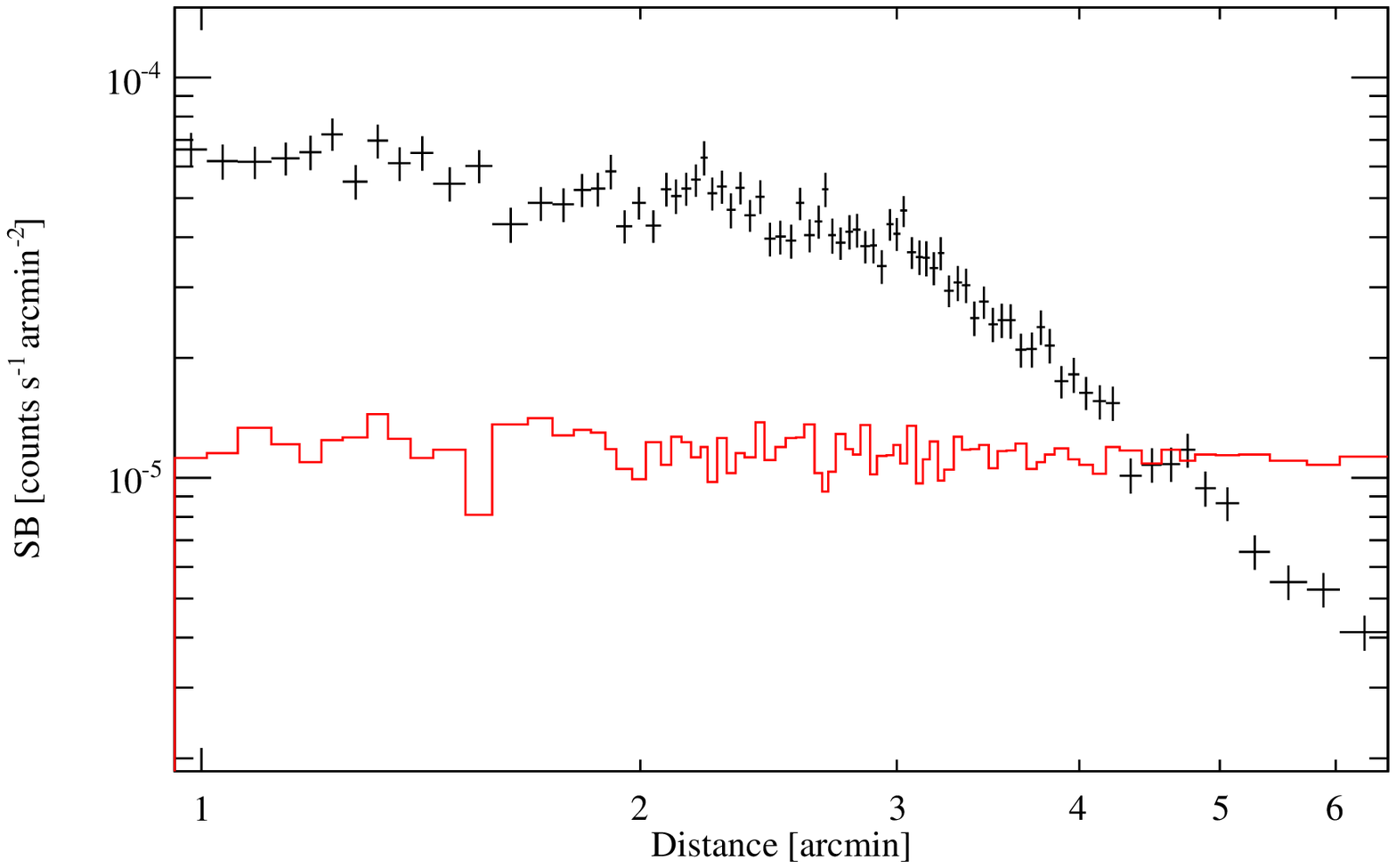}
  \includegraphics[width=\columnwidth]{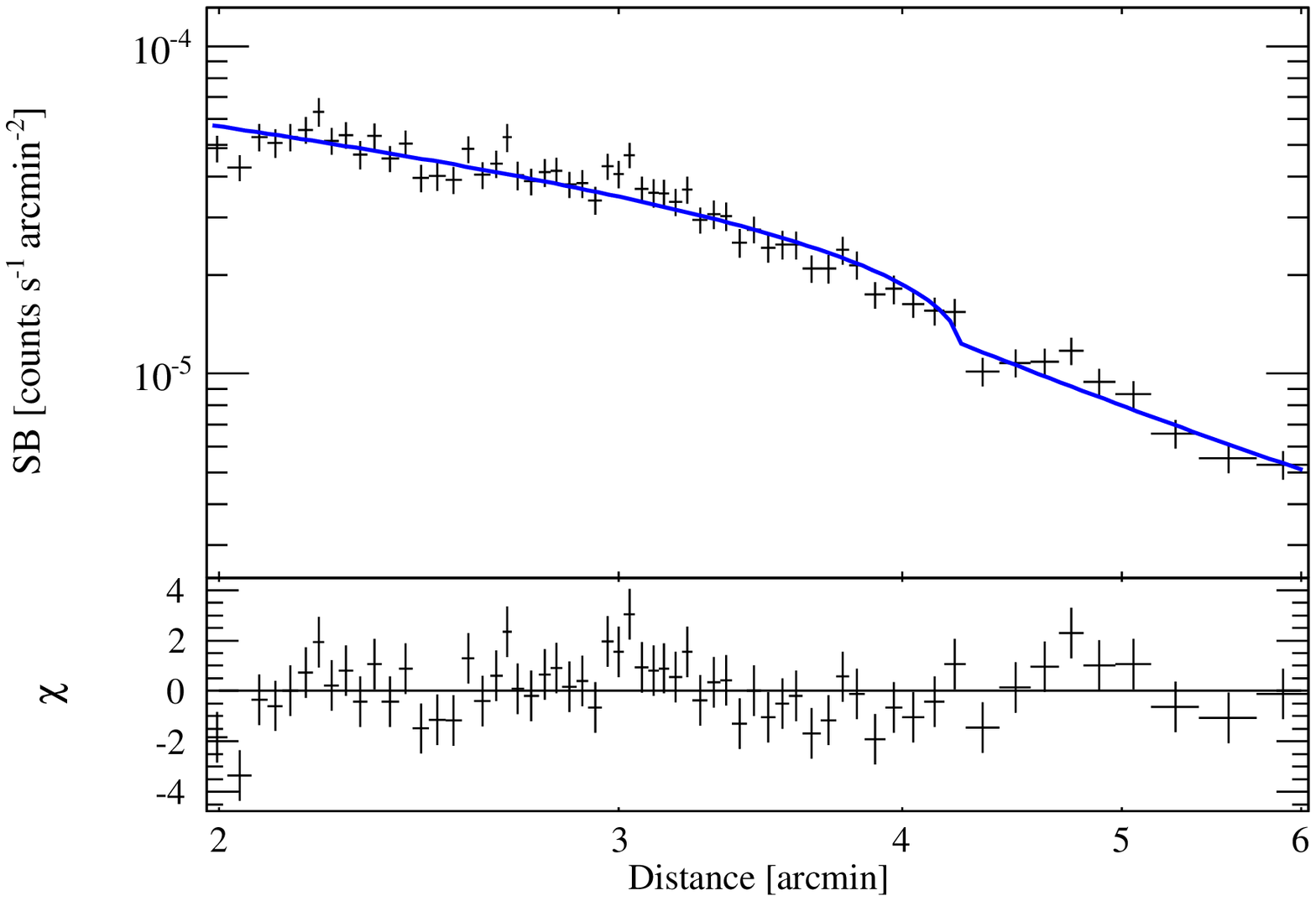}
  \caption{Surface brightness profiles extracted from sector S4 in Figure \ref{fig:sectors}. The profiles were binned to a uniform SNR of 7. The top panel displays the full surface brightness profile outside a radius of 1 arcmin, with the instrumental background (shown in red) subtracted. Below, we show part of this profile, fitted with a broken power-law model (shown in blue). The discontinuity correspond to $\mathcal{M} = 1.26_{-0.07}^{+0.08}$.}
  \label{fig:e-sx}
\end{figure}

The western and eastern surface brightness profiles are presented in Figures \ref{fig:w-sx} and \ref{fig:e-sx}. Both profiles display surface brightness edges. We fitted these edges with broken power-law models, and summarized the best-fit parameters in Table \ref{tab:sxfits}. The W and E edges correspond to Mach numbers of $1.25_{-0.06}^{+0.07}$ and $1.26_{-0.07}^{+0.08}$, respectively. To the W, there is a hint of an additional discontinuity around 7 arcmin, yet this could not be satisfactorily fitted with our model because the number of bins in the putative preshock region is very small.

We evaluated the effect of a $1\sigma$ background surface brightness uncertainty on the Mach numbers determined at each of the surface brightness discontinuities. The background level was fixed to its best-fit minus $1\sigma$ and plus $1\sigma$ values, and all profiles were refitted for the new background level. The Mach numbers did not change significantly. 

To further test the robustness of our results, we employed a complementary analysis of the background, in which the background level in each sector was calculated from the outer bins of the individual surface brightness profiles. Because the statistics in the outer bins are rather poor, we used Cash statistics \citep{Cash1979}. Therefore, the background was not subtracted from the profiles, and the sky background was simply modelled as a constant that was added to the instrumental background surface brightness profile. Before fitting, we binned the profiles to a minimum of 3 counts/bin. The background levels for the N, S, W, and E sectors were, respectively, $(1.25\pm 0.09)\times 10^{-6}$, $(1.40\pm 0.19)\times 10^{-6}$, $(1.74\pm 0.11)\times 10^{-6}$, and $(4.97\pm 1.20)\times 10^{-7}$ cts/s/arcmin$^2$. The surface brightness profiles in all sectors were refitted using Cash statistics with a model consisting of the sum of a power-law density model projected along the line of sight, a constant representing the sky background, and the instrumental background surface brightness profile. The sky background was fixed in the fits to the corresponding sky background levels determined previously. All other fit parameters were free. The results of the fits are summarized in Table \ref{tab:sxfits-cash}. The Mach numbers derived from the density jumps are generally consistent with the values in Table \ref{tab:sxfits}; the largest difference is seen for the S sector, but even here the results are consistent within $2\sigma$.

We also attempted to fit the profiles with a simple power-law in the same radii ranges used for the broken power-law fits. These fits were rather poor, as shown in Table \ref{tab:sxfits}.

\begin{figure*}
  \includegraphics[width=\textwidth]{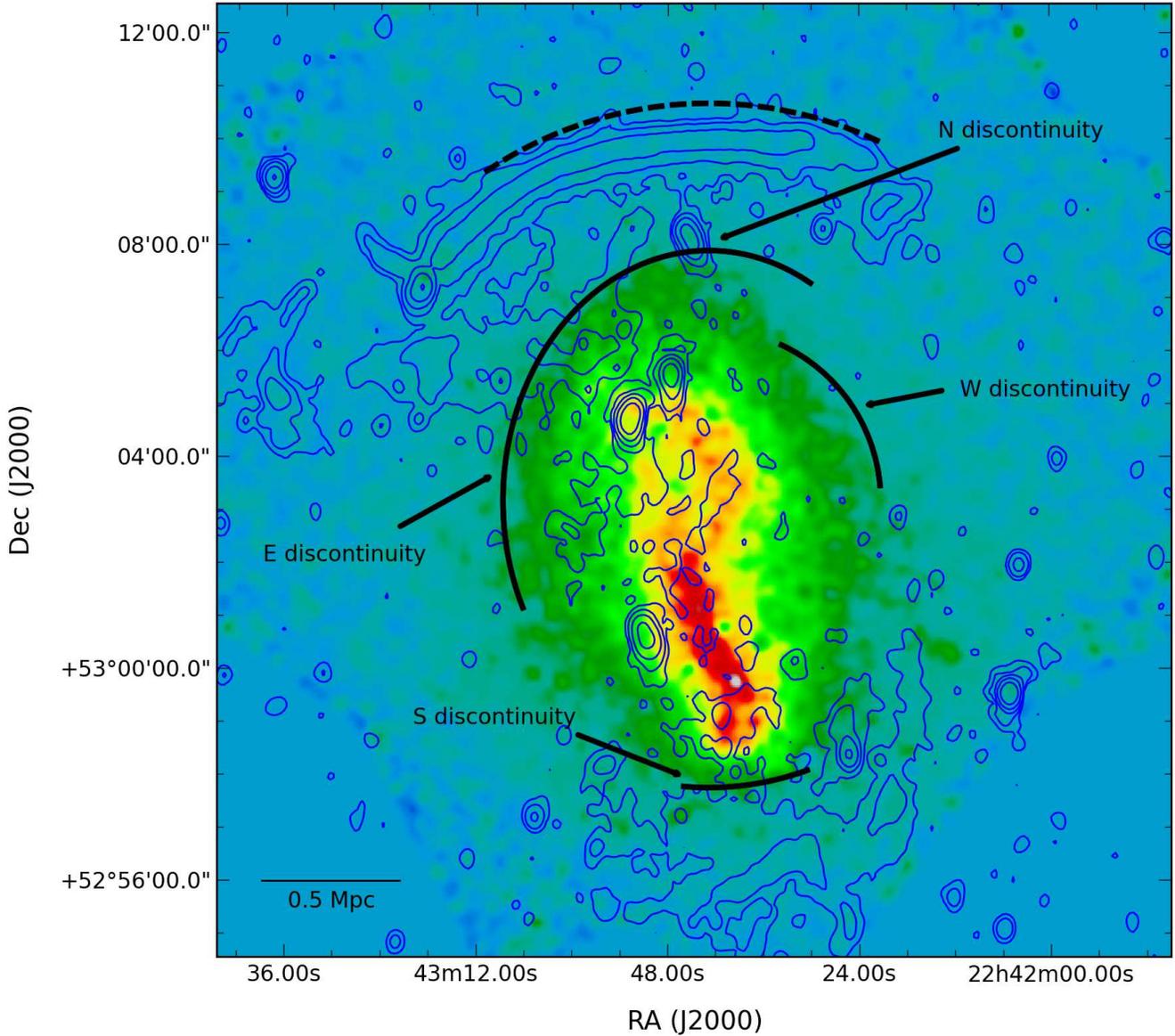}
  \caption{Same as Figure \ref{fig:sxmap-radio}, with overlaid arcs showing the location of the discontinuities identified in the surface brightness profiles. The discontinuities are labeled. The dashed line marks the radius at which we detect a hint of a density jump in the surface brightness profile, and also the outer edge of the northern radio relic. Because a relic is present at this position, the northern discontinuity is more likely than the very weak one detected W of the relic; therefore, the latter is not shown.}
  \label{fig:shocks}
\end{figure*}

\begin{figure}
  \includegraphics[width=\columnwidth]{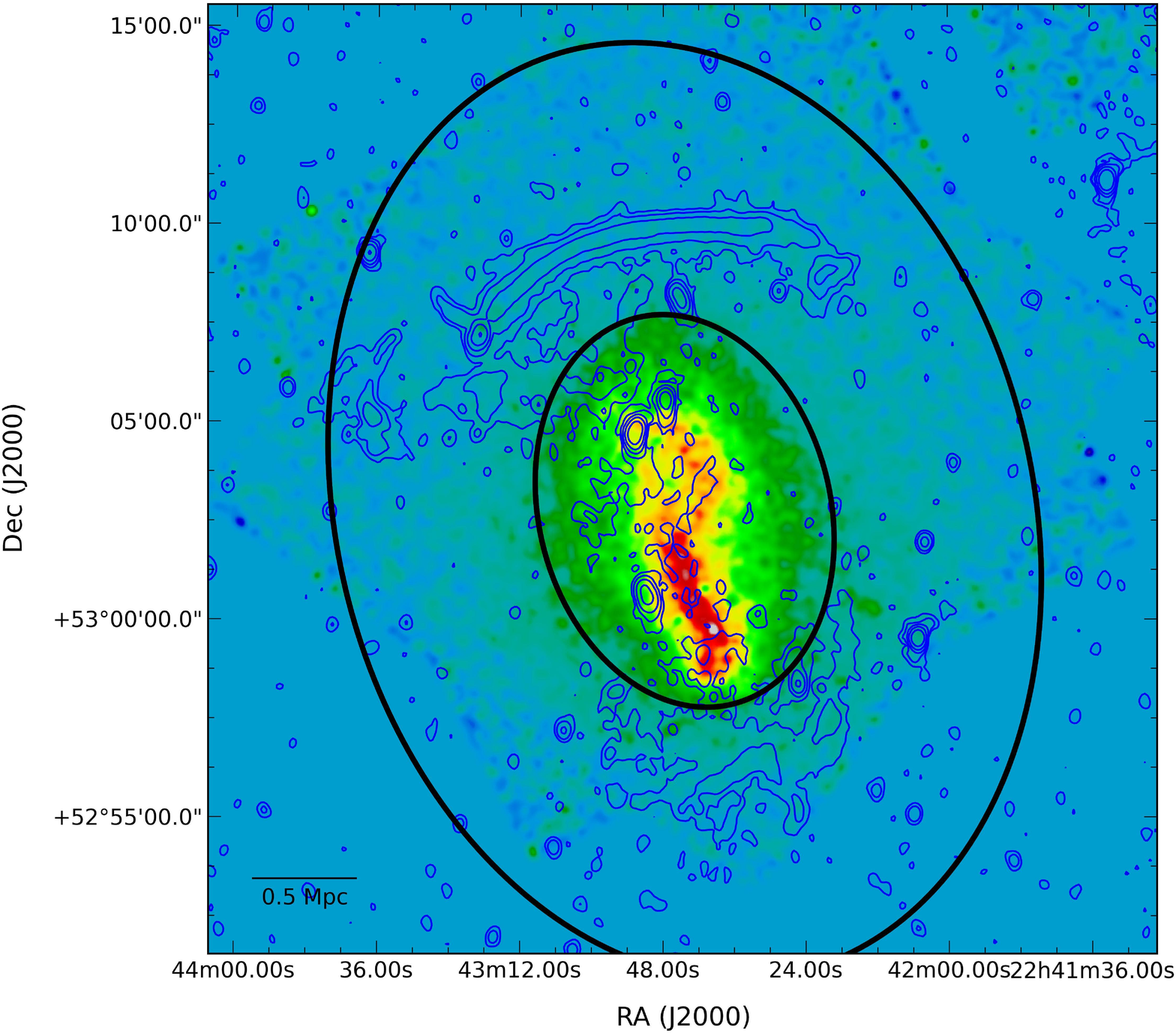}
  \caption{Same as Figure \ref{fig:sxmap-radio}, with two overlaid ellipses showing the region used for extracting the surface brightness profile in Figure \ref{fig:sxell} and an additional inner ellipse that represents an approximation to the inner shocks' shapes. The ellipse was chosen to follow the inner shocks.}
  \label{fig:ellipse}
\end{figure}

\begin{figure}
  \includegraphics[width=\columnwidth]{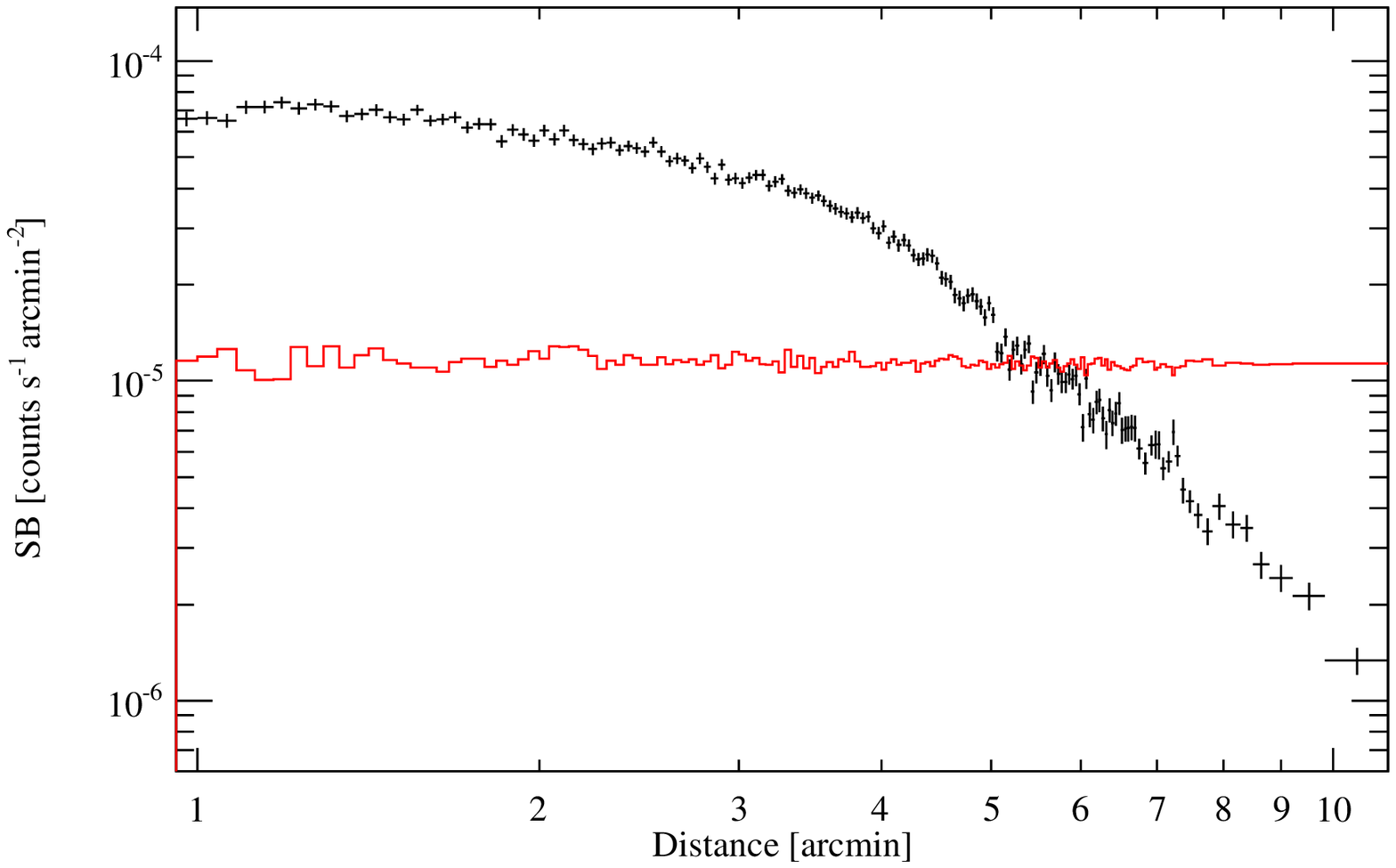}
  \includegraphics[width=\columnwidth]{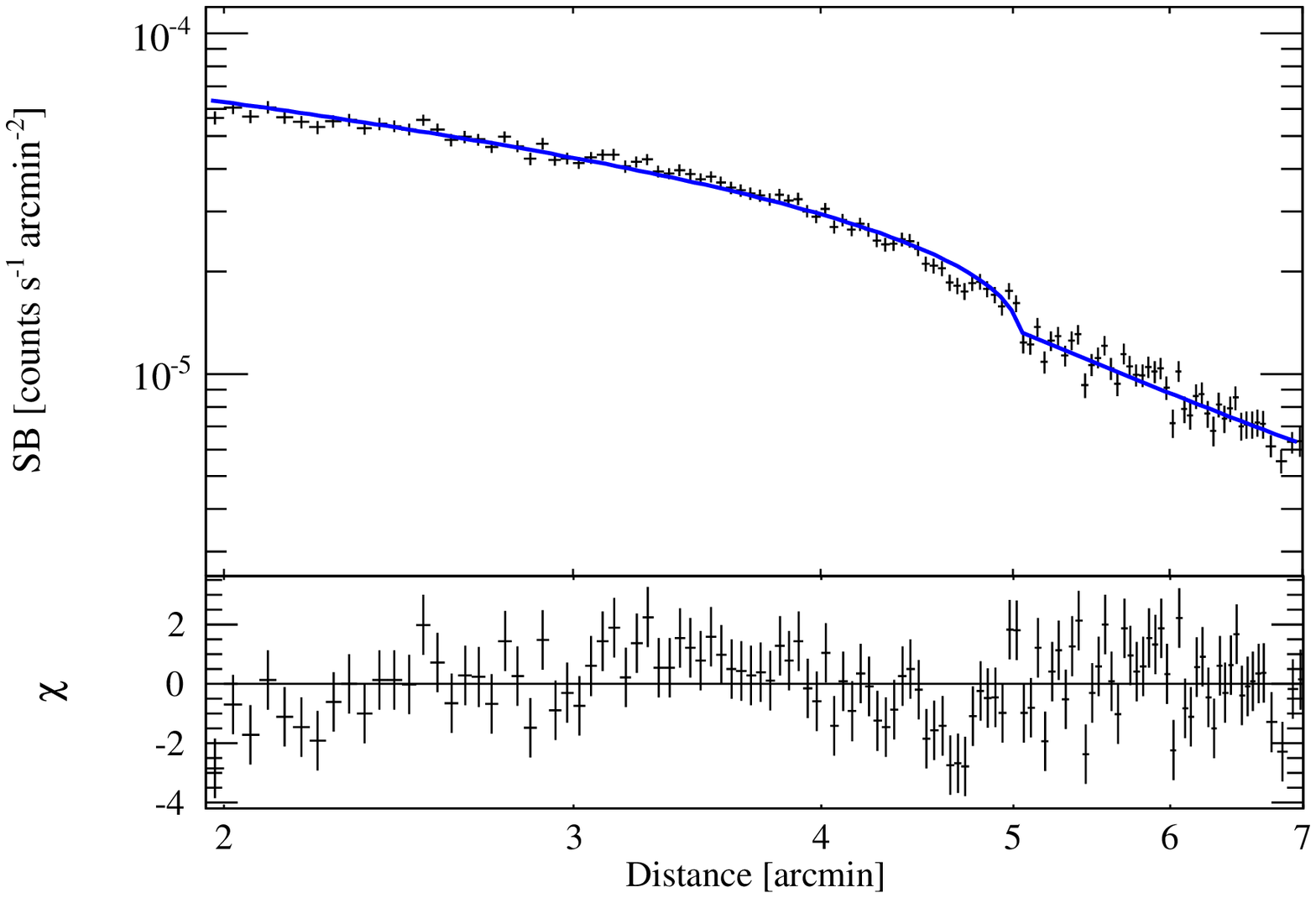}
  \caption{Surface brightness profile in elliptical annuli around the cluster centre. The profile was binned to a uniform SNR of 10. The x-axis shows the distance from the centre, measured along the major axis. The figure at the top displays the full surface brightness profile outside a radius of 1 arcmin, with the instrumental background (shown in red) subtracted. At the bottom we show part of this profile, fitted with a broken power-law model (shown in blue). The discontinuity correspond to a Mach number of $1.27\pm 0.03$. We note that the presence of this discontinuity does not necessarily prove that a shock goes all around the cluster; the ellipse was specifically chosen to follow all the four surface brightness discontinuities.}
  \label{fig:sxell}
\end{figure}

In Figure \ref{fig:shocks}, we show again the surface brightness map, with overlaid arcs marking the positions of the surface brightness jumps. 

To constrain the inner density jumps better, we extracted a surface brightness profile in elliptical annuli around the cluster centre. The chosen ellipse follows the inner surface brightness discontinuities. It has a major axis of 12 arcmin, a minor axis of 8.9 arcmin, and an angle between the major axis and the right ascension axis of 104 degrees. The ellipse is shown in Figure \ref{fig:ellipse}, while the surface brightness profile, binned to a uniform SNR of 10, is shown in Figure \ref{fig:sxell}. The profile was fitted around the discontinuity at $\sim 5$ arcmin (measured along the major axis) with a broken power-law that had all the parameters free, plus a fixed constant describing the average sky background. The best-fit model is summarized in Table \ref{tab:sxfits}. The shock compression corresponds to a Mach number of $1.27\pm 0.03$. We note that while this discontinuity proves the existence of the density jumps, it does not prove that the shock goes uninterrupted all around the cluster. In addition, the assumption of spherical symmetry which stands behind the density model is clearly violated.

\begin{table*}
  \caption{Best-fit surface brightness parameters for the models fitted to the profiles shown in Figures \ref{fig:n-sx}, \ref{fig:s-sx}, \ref{fig:e-sx}, \ref{fig:w-sx}, and \ref{fig:sxell} using $\chi^2$ statistics. For comparison, we also list the $\chi^{2}/{\rm d.o.f.}$ obtained when a simple power-law model was fitted in the same radius range. }
  \label{tab:sxfits}
  \smallskip
  \begin{threeparttable}
    \begin{tabular}{lccccccc}
      \hline
	    & $\alpha_1$ & $\alpha_2$ & $r_{\rm sh}$\tnote{a} & $C$ & $\mathcal{M}$ & $\chi^2/{\rm d.o.f.}$ & $\chi^2/{\rm d.o.f.}_{\rm PL}$\tnote{b}  \\
      \hline
	N & $0.61_{-0.13}^{+0.12}$ & $2.26_{-0.27}^{+0.24}$ & $4.69_{-0.06}^{+0.05}$ & $1.46_{-0.13}^{+0.17}$ & $1.31_{-0.09}^{+0.12}$ & $1.06$ & $3.12$ \\ 
	S & $1.09\pm 0.18$ & $2.81\pm 0.24$ & $5.42\pm 0.02$ & $1.62_{-0.11}^{+0.12}$ & $1.43_{-0.08}^{+0.09}$ & $1.41$ & $3.39$ \\
	E & $0.78\pm 0.07$ & $2.01\pm 0.18$ & $4.24\pm 0.24$ & $1.39_{-0.10}^{+0.11}$ & $1.26_{-0.07}^{+0.08}$ & $1.47$ & $4.86$ \\
	W & $0.79_{-0.06}^{+0.10}$ & $1.53_{-0.17}^{+0.09}$ & $3.24_{-0.15}^{+0.19}$ & $1.37_{-0.09}^{+0.10}$ & $1.25_{-0.06}^{+0.07}$ & $0.84$ & $2.59$ \\
	Ellipse & $0.69\pm 0.02$ & $1.83\pm 0.07$ & $5.04_{-0.02}^{+0.03}$ & $1.40\pm 0.04$ & $1.27\pm 0.03$ & $1.14$ & $5.72$ \\
      \hline
    \end{tabular}
    \begin{tablenotes}
      \item[\hspace{0.3cm}a] In units of arcmin.
      \item[\hspace{0.3cm}b] $\chi^2/{\rm d.o.f.}_{\rm PL}$ obtained with a power-law model.
    \end{tablenotes}
  \end{threeparttable}
\end{table*}

\begin{table*}
  \caption{Best-fit surface brightness parameters for the models fitted to the profiles shown in Figures \ref{fig:n-sx}, \ref{fig:s-sx}, \ref{fig:e-sx}, \ref{fig:w-sx}, and \ref{fig:sxell} using Cash statistics. For comparison, we also list the Mach numbers obtained in the $\chi^2$ fits.}
  \label{tab:sxfits-cash}
  \smallskip
  \begin{threeparttable}
    \begin{tabular}{lcccccc}
      \hline
	    & $\alpha_1$ & $\alpha_2$ & $r_{\rm sh}$\tnote{a} & $C$ & $\mathcal{M}_{\rm Cash}$ & $\mathcal{M}_{\chi^2}$  \\
      \hline
	N & $0.61\pm 0.09$ & $2.19_{-0.20}^{+0.21}$ & $4.71\pm 0.02$ & $1.51_{-0.11}^{+0.13}$ & $1.35_{-0.08}^{+0.09}$ & $1.31_{-0.09}^{+0.12}$ \\ 
	S & $0.89\pm 0.19$ & $2.84\pm 0.12$ & $5.30\pm 0.09$ & $1.32\pm 0.06$ & $1.21\pm 0.04$ & $1.43_{-0.08}^{+0.09}$ \\
	E & $0.74\pm 0.07$ & $1.93\pm 0.14$ & $4.16_{-0.07}^{+0.02}$ & $1.34\pm 0.08$ & $1.23\pm 0.05$ & $1.26_{-0.07}^{+0.08}$ \\
	W & $0.81\pm 0.05$ & $1.63\pm 0.08$ & $3.22_{-0.12}^{+0.16}$ & $1.35_{-0.08}^{+0.09}$ & $1.23\pm 0.06$ & $1.25_{-0.06}^{+0.07}$ \\
      \hline
    \end{tabular}
    \begin{tablenotes}
      \item[\hspace{0.3cm}a] In units of arcmin.
    \end{tablenotes}
  \end{threeparttable}
\end{table*}

\section{Spatially-resolved spectroscopy}
\label{s:spectroscopy}

At shock fronts, the temperature on the inner side of the density jump is higher than on the outer side. The opposite is true for cold fronts. Therefore, measurements of the temperature on both sides of a density discontinuity are required to undoubtedly reveal the nature of the jump. Using archival \suzaku\ observations, we examined the temperature in the inner and outer regions of the surface brightness discontinuities determined in the previous section. The \suzaku\ datasets analysed here are the same as those previously presented by \citet{Akamatsu2013}.

\subsection{\suzaku\ data reduction}

CIZA J2242.8+5301 was observed for 123 ks with Suzaku, on 2011 July 28. For the purpose of the background analysis, an offset observation was performed on 2011 July 30, for 56 ks. The two data sets -- ObsIDs 806001010 and 806002010 -- are publicly available in the Suzaku database (PI: Kawahara).

The data reduction was done with HEAsoft v6.12 and the calibration files released on 2013 March 5. Each event file was fully reprocessed and screened using the default screening parameters\footnote{http://heasarc.nasa.gov/docs/suzaku/processing/criteria\_xis.html}, with the exception of the parameters constraining the elevation of the source (ELV) and the geomagnetic cut-off rigidity (COR), which were set to ${\rm ELV} > 10$ degrees and ${\rm COR} > 6$ GeV. For each observation and detector, we then merged the event files in the 3x3 and 5x5 editing modes. Point sources were detected with the Chandra Interactive Analysis of Observations (CIAO) task \textsc{wavdetect} in the energy band $0.5-8$ keV down to a flux limit of $\sim 4\times 10^{-14} \rm{erg\,\,s^{-1}\,\,cm^{-2}}$. This is the lowest point source flux in the offset observation. For consistency, we only considered point sources in the on-target observation down to the same flux limit obtained from the background pointing. All point sources were excluded using circular regions with 2 arcmin radii. Furthermore, we excluded the corner regions illuminated by the $^{55}$Fe calibration sources, and the columns on the XIS0 detector that were affected by a micro-meteorite hit in June 2009.

\subsection{Spectral analysis}

\begin{table}
  \caption{Best-fit background model. }
  \label{tab:bkg}
  \smallskip
  
  \begin{threeparttable}
    \begin{tabular}{lccc}
      \hline
	    & $T$\tnote{a} & $\Gamma$\tnote{b} & $\mathcal{N}$\tnote{c} \\
      \hline
	LHB &  $0.08$\tnote{d} & -- & $1.74_{-0.17}^{+0.18} \times 10^{-5}$ \\ 
	GH  &  $0.30\pm 0.01$ & -- & $(3.49\pm 0.27) \times 10^{-6}$ \\
	CXB & -- & $1.41$\tnote{d} & $(7.31\pm 0.28)\times 10^{-7}$ \\
      \hline
    \end{tabular}
    \begin{tablenotes}
      \item[\hspace{0.3cm}a] Temperature in units of keV.
      \item[\hspace{0.3cm}b] Spectral index.
      \item[\hspace{0.3cm}c] Normalization in default \textsc{XSpec} units per $1/400\pi$.
      \item[\hspace{0.3cm}d] Frozen parameter.
    \end{tablenotes}
  \end{threeparttable}
\end{table}

XIS0, XIS1, and XIS3 spectra were extracted from the whole FOV of the Suzaku offset observation, and from partial annuli behind and ahead of the N, E, and W surface brightness discontinuities detected in Section \ref{s:analysis}.~\footnote{The S region is not covered by the archived Suzaku datasets. Results from new observations covering the southern region will be published by Akamatsu et al. (in prep.).} We also extracted spectra on both sides of the outer N discontinuity that is only weakly seen in the surface brightness profile. This discontinuity, if true, would be associated with the shock traced by the relic. For each spectrum, we generated RMF and ARF files. We also extracted corresponding instrumental/non-X-ray background (NXB) spectra from the NXB event files obtained after June 2011, with the increased charge injection level of 6 keV. The spectra were NXB-subtracted and fitted with a model that consisted of X-ray background and ICM emission. 

The X-ray background was modelled as the sum of two thermal and one non-thermal emission components, similar to the approach used by \citet{Akamatsu2013}. The thermal components represent emission from the Local Hot Bubble (LHB) and the Galactic Halo (GH), while the non-thermal component describes emission from unresolved point sources (cosmic X-ray background; CXB). For the thermal components, the redshifts and metallicities were fixed to 0 and 1, respectively. 

The ICM emission was described by a single temperature model with free temperature and normalisation, and fixed redshift ($z=0.1921$) and metallicity ($Z=0.2Z_{\sun}$).

To correctly propagate the errors, all the spectra were fitted simultaneously. The results of the fit are summarized in Tables \ref{tab:bkg} and \ref{tab:icm}. The fit had $\chi^2/{\rm d.o.f.}=0.98$.

Across the northern relic, the temperature decreases from $9.57_{-1.12}^{+1.25}$ to $3.35_{-0.72}^{+1.13}$ keV, which indicates a shock front of Mach number $2.54_{-0.43}^{+0.64}$, consistent with that derived by \citet{Akamatsu2013}. However, across the inner northern discontinuity, the temperatures are roughly the same. West and east of the merger axis, the temperatures are also very similar on both sides of the surface brightness discontinuities. The Mach numbers associated with the temperature ratios on the two sides of the density discontinuities are consistent, within $\sim 2\sigma$, with the low Mach numbers derived from the shock compression at the inner density jumps. Nevertheless, no clear evidence of temperature jumps is found at the inner surface brightness discontinuities. In the next section, we discuss possible explanations.

\begin{table*}
  \caption{Best-fit ICM parameters for the spectra extracted from the inner (2) and outer (1) side of the surface brightness discontinuities discussed in Section \ref{s:analysis}. We derived the Mach number corresponding to the temperature ratio $T_2/T_1$ and, for ease of comparison, we list again the Mach numbers derived from the density jumps.}
  \label{tab:icm}
  \smallskip
 
  \begin{threeparttable}
    \begin{tabular}{lcccccc}
      \hline
	    & $T_1$\tnote{a} & $T_2$\tnote{a} & $\mathcal{N}_1$\tnote{b} & $\mathcal{N}_2$\tnote{b} & $\mathcal{M}_{\rm T}$\tnote{c} & $\mathcal{M}$\tnote{d} \\
      \hline
	N (relic)\tnote{e} & $3.35_{-0.72}^{+1.13}$ & $9.57_{-1.12}^{+1.25}$ &  $3.03_{-0.49}^{+0.54}\times 10^{-6}$ & $(1.77\pm 0.07)\times 10^{-5}$ & $2.54_{-0.43}^{+0.64}$ & -- \\
	N (inner shock) & $8.58_{-0.96}^{+1.18}$ & $7.99_{-0.52}^{+0.53}$ & $1.96_{-0.09}^{+0.10} \times 10^{-5}$ & $(6.07\pm 0.18) \times 10^{-5}$ & $0.93_{-0.12}^{+0.14}$ & $1.31_{-0.09}^{+0.12}$ \\
	W & $6.40_{-0.67}^{+0.77}$ & $6.59_{-0.42}^{+0.66}$ & $(2.43\pm 0.14) \times 10^{-5}$ & $7.50_{-0.28}^{+0.29}\times 10^{-5}$ & $1.03_{-0.13}^{+0.16}$ & $1.25_{-0.06}^{+0.07}$ \\
	E & $8.38_{-0.70}^{+1.07}$ & $9.20_{-0.66}^{+0.67}$ & $(2.71\pm 0.10) \times 10^{-5}$ & $(9.73\pm 0.22)\times 10^{-5}$ & $1.10_{-0.12}^{+0.17}$ & $1.26_{-0.07}^{+0.08}$ \\
      \hline
    \end{tabular}
 
    \begin{tablenotes}
      \item[\hspace{0.3cm}a] Temperature in units of keV.
      \item[\hspace{0.3cm}b] Normalization in default \textsc{XSpec} units per $1/400\pi$.
      \item[\hspace{0.3cm}c] Mach number derived from the temperature ratio $T_2/T_1$.
      \item[\hspace{0.3cm}d] Mach number derived from the density jump (also listed in Table \ref{tab:sxfits}).
      \item[\hspace{0.3cm}e] \citet{Akamatsu2013} found a northern postshock temperature of $8.3\pm 0.8$ keV and a preshock temperature of $2.1\pm 0.4$ keV. However, they did not exclude any point sources from the FOV.
    \end{tablenotes}
  \end{threeparttable}
\end{table*}

\section{Discussion and conclusions}
\label{s:discussion}

Analysing the surface brightness profiles along directions on- and off-relic, we identified density discontinuities in all directions from the cluster centre. These discontinuities correspond to very weak Mach numbers of $\sim 1.3$. The only clear temperature jump is detected across the northern relic; it corresponds to a Mach number of $2.54_{-0.43}^{+0.64}$. However, the temperatures throughout the cluster are very high, larger than 5 keV out to distances as large as 1.5 Mpc from the cluster centre \citep{Ogrean2013b}. These high temperatures are difficult to constrain given the useful energy range of \suzaku\ (as well as \xmm\ and \chandra), roughly $0.5-7$ keV. The large statistical uncertainties on the measured temperatures make it impossible to confirm in temperature the Mach numbers calculated from the surface brightness discontinuities. Yet, small temperature jumps associated with Mach numbers of $\sim 1.3$ cannot be excluded.

The density discontinuities identified in the \chandra\ observations of CIZA J2242.8+5301 raise several questions related to particle acceleration at cluster merger shocks:
\begin{enumerate}
  \item How does the temperature jump detected across the northern relic compare with the radio-predicted Mach number?
  \item What is the nature of the inner discontinuities?
  \item Why is there no radio emission associated with the detected density edges?
  \item What is the merger scenario that triggered multiple sequential discontinuities in the ICM?
\end{enumerate}
We address each of these questions below.

\subsection{The shock at the N relic}

The outer northern discontinuity in surface brightness is only weakly detected. A density jump was expected here, as it would trace the edge of the ``Sausage'' relic and a temperature jump has already been detected near its location by \citet{Akamatsu2013}. We also detect a temperature jump. However, while the Mach number inferred from the temperature jump is relatively small, $\mathcal{M} = 2.54_{-0.43}^{+0.64}$, the Mach number predicted based on the radio spectral index under the assumptions of diffusive shock acceleration in the test-particle regime \citep{Drury1983} is much larger, $\mathcal{M}=4.6\pm 1.1$ \citep{vanWeeren2010}. The same discrepancy between the radio-predicted and X-ray-derived Mach numbers was observed at the northern relic in 1RXS J0603.3+4214 \citep{Ogrean2013d}. Numerical simulations are required to clarify the reasons for the discrepancy. One possibility is that the Mach number varies across the shock front \citep{Skillman2013}, and the synchrotron emission is more sensitive to high Mach number shocks \citep{HoeftBrueggen2007}. Alternatively, the shock does not only accelerate particles from the thermal pool, but re-accelerates a pre-existing cosmic ray particle population. This scenario has been simulated by \citet{Kang2012}, who showed that the northern relic can be reproduced either by a shock of Mach number $\sim 4$, or by a weaker shock of Mach number $\sim 2$. Other possible explanations are projection effects, underestimation of the postshock temperature due to averaging in a wide partial annulus (width $\sim 450$ kpc), or oblique, rather than parallel, shocks \citep{Kirk1989}.

\subsection{The nature of the inner density discontinuities}

The nature of the inner density discontinuities is difficult to explain in the absence of temperature jumps. On one hand, it seems rather unlikely that the northern discontinuity is a cold front, given its large distance from the centre. At $\approx 1.5$ Mpc from the centre, a cold front in the northern region of CIZA J2242.8+5301 would be the most distant cold front ever detected. At the moment, the most known distant cold front was found in the cluster Abell 2142, at 1 Mpc from the centre \citep{Rossetti2013}. Moreover, while the \xmm\ temperature in the direction of the northern relic revealed a hint of a temperature increase on the outer side of the discontinuity \citep{Ogrean2013a}, this result is not confirmed by \emph{Suzaku}. The temperature on both sides of the northern discontinuity is $\sim 8-9$ keV, which would also mean that if this discontinuity was associated with a cold front, it would be the hottest of all known cold fronts. Nevertheless, while there are indications that at least the northern discontinuity is more likely a shock front, only more precise temperature measurements can confirm this supposition as well as identify the nature of the other inner density jumps.

\subsection{No radio emission at the inner density discontinuities}

Interestingly, no diffuse radio emission is detected at either of the inner surface brightness discontinuities. If these discontinuities are associated with shock fronts, then the situation is similar to that in Abell 2146 \citep{Russell2010}. However, unlike in Abell 2146, the shocks are not present only along the merger axis, but also west and east of it. More importantly from a particle acceleration viewpoint, while the shock front in Abell 2146 has a Mach number of $\approx 2$, the inner density discontinuities in CIZA J2242.8+5301 appear to be associated with much weaker shocks. Indeed, DSA of thermal particles at very weak shocks does not efficiently accelerate electrons to cosmic ray energies \citep{Kang2007}. Moreover, for a $\mathcal{M}=1.3$ shock, DSA of thermal particles in the test-particle regime would imply an extremely steep spectral index of $\approx 4$, making a relic, if present, too faint in the radio band.

Another possibility is that radio relics require a pre-existing CR particle population \citep[e.g.,][]{Kang2007,Kang2012}. In that case, the lack of radio emission at the inner, W, and E shocks suggests that no pre-existing CR population existed at the shock locations previous to the shock passage. However, even if a pre-existing CR population was present, re-acceleration would boost the electron spectrum only by $3C/[C+2-\delta (C-1)] \approx 2$ for a shock of Mach number $1.3$ and a slope of the pre-existing particle population spectrum $\delta=2-3$ \citep{Markevitch2005}.

If the discontinuities are associated with cold fronts rather than with shocks, then radio emission is naturally not expected.

\subsection{Sequential shocks in the ICM}

\begin{figure}
  \includegraphics[width=\columnwidth]{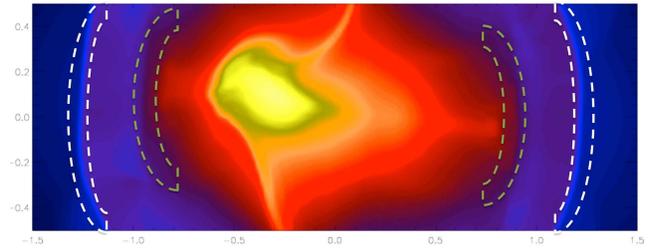}
  \caption{Logarithm of the gas density in a cut through the plane that contains the two cluster centres. The location of the two sets of discontinuities are indicated by the dashed-line regions.}
  \label{fig:sim}
\end{figure}

If the inner discontinuities are shock fronts, then it is for the first time that multiple sequential merger shocks are observed in a galaxy cluster. In numerical simulations, multiple consecutive shocks have been seen in the accretion region of clusters \citep[e.g.,][]{Vazza2009} and close to an active AGN \citep[][]{Brueggen2007}, but not at moderate distances ($\sim 1$ Mpc) from the cluster centre. However, secondary shocks in a binary cluster merger could stem (i) from a second core passage of the DM cores, or (ii) from the violent relaxation of the newly merged halo. In the former case, the discontinuities are strongest along the merger axis and in the latter case, the shocks can be radial with a weaker variation with angle from the merger axis. In order to better understand how multiple shocks can be produced, we have performed very simple dark matter (DM) + hydrodynamic simulations of binary cluster mergers using the adaptive-mesh refinement code {\sc flash}. For more details on the initial conditions and the physics employed in these simulations see \citet{Brueggen2012}. In these idealised simulations of nearly equal mass mergers, we set two clusters in virial and hydrostatic equilibrium on a collision course with a small impact parameter to destroy axial symmetry. Note that these simple simulations were set up to study the interaction between the gas and the DM and were not tuned to reproduce conditions in CIZA J2242.8+5301.  In Fig. \ref{fig:sim} we show the logarithm of the gas density in a cut through the plane that contains the two cluster centres. The outermost shock is produced by the impact of the two gas spheres, which lead to a region of high compression and an ellipsoidal shock waves that run outwards. Meanwhile, the DM halos get compressed and tidal tails develop at the far sides of the DM halos. When the DM cores separate again after core passage, they go through violent relaxation as the total gravitational potential varies rapidly and the DM cores expand swiftly (on the violent relaxation time scale which is the timescale on which the gravitational potential changes). This rapid expansion can cause another set of shocks (usually weaker) through the gas, and this is seen as the inner discontinuity. Curiously, in the literature on cluster mergers, this is almost never mentioned. Violent relaxation is mentioned but not its impact on the ICM, presumable because it is dynamically not important and it has never been observed before. The possibility of two shocks is mentioned, however, in \citet{Birnboim2010}, who studied merging shock fronts in the context of cold fronts in galaxy clusters.

Finally, the presence of multiple putative shocks could also suggest a more complex merger geometry than a simple binary, head-on collision, and complex subsequent processes within the ICM. We leave the investigation of the origin of these possible shock structures to future papers.


\section*{Acknowledgments}

We thank the referee for constructive comments that significantly improved the manuscript. GAO thanks M. Markevitch for his help with the \chandra\ background analysis. GAO, MB, and MH acknowledge support by the research group FOR 1254 funded by the Deutsche Forschungsgemeinschaft (DFG). RJvW acknowledges support provided by NASA through Einstein Postdoctoral Fellowship grant number PF2-130104 awarded by the Chandra X-ray Center, which is operated by the Smithsonian Astrophysical Observatory for NASA under contract NAS8-03060.  MB acknowledges allocations 5056 and 5984 on supercomputers
at the NIC of the Forschungszentrum J\"{u}lich. The scientific results reported in this article are based on observations made by the Chandra X-ray Observatory.

\bibliographystyle{mn2e}
\bibliography{bibliography}

\begin{thebibliography}{}

\bibitem[\protect\citeauthoryear{{Akamatsu} \& {Kawahara}}{{Akamatsu} \&
  {Kawahara}}{2013}]{Akamatsu2013}
{Akamatsu} H.,  {Kawahara} H.,  2013, \pasj, 65, 16

\bibitem[\protect\citeauthoryear{{Birnboim}, {Keshet} \&
  {Hernquist}}{{Birnboim} et~al.}{2010}]{Birnboim2010}
{Birnboim} Y.,  {Keshet} U.,    {Hernquist} L.,  2010, \mnras, 408, 199

\bibitem[\protect\citeauthoryear{{Bourdin}, {Mazzotta}, {Markevitch},
  {Giacintucci} \& {Brunetti}}{{Bourdin} et~al.}{2013}]{Bourdin2013}
{Bourdin} H.,  {Mazzotta} P.,  {Markevitch} M.,  {Giacintucci} S.,
  {Brunetti} G.,  2013, \apj, 764, 82

\bibitem[\protect\citeauthoryear{{Br{\"u}ggen}, {Heinz}, {Roediger},
  {Ruszkowski} \& {Simionescu}}{{Br{\"u}ggen} et~al.}{2007}]{Brueggen2007}
{Br{\"u}ggen} M.,  {Heinz} S.,  {Roediger} E.,  {Ruszkowski} M.,
  {Simionescu} A.,  2007, \mnras, 380, L67

\bibitem[\protect\citeauthoryear{{Br{\"u}ggen}, {van Weeren} \&
  {R{\"o}ttgering}}{{Br{\"u}ggen} et~al.}{2012}]{Brueggen2012}
{Br{\"u}ggen} M.,  {van Weeren} R.~J.,    {R{\"o}ttgering} H.~J.~A.,  2012,
  \mnras, 425, L76

\bibitem[\protect\citeauthoryear{{Brunetti} \& {Jones}}{{Brunetti} \&
  {Jones}}{2014}]{Brunetti2014}
{Brunetti} G.,  {Jones} T.~W.,  2014, ArXiv e-prints

\bibitem[\protect\citeauthoryear{{Cash}}{{Cash}}{1979}]{Cash1979}
{Cash} W.,  1979, \apj, 228, 939

\bibitem[\protect\citeauthoryear{{Drury}}{{Drury}}{1983}]{Drury1983}
{Drury} L.~O.,  1983, Reports on Progress in Physics, 46, 973

\bibitem[\protect\citeauthoryear{{Eckert}, {Molendi} \& {Paltani}}{{Eckert}
  et~al.}{2011}]{Eckert2011b}
{Eckert} D.,  {Molendi} S.,    {Paltani} S.,  2011, \aap, 526, A79

\bibitem[\protect\citeauthoryear{{Feretti}, {Giovannini}, {Govoni} \&
  {Murgia}}{{Feretti} et~al.}{2012}]{Feretti2012}
{Feretti} L.,  {Giovannini} G.,  {Govoni} F.,    {Murgia} M.,  2012, \aapr, 20,
  54

\bibitem[\protect\citeauthoryear{{Finoguenov}, {Sarazin}, {Nakazawa}, {Wik} \&
  {Clarke}}{{Finoguenov} et~al.}{2010}]{Finoguenov2010}
{Finoguenov} A.,  {Sarazin} C.~L.,  {Nakazawa} K.,  {Wik} D.~R.,    {Clarke}
  T.~E.,  2010, \apj, 715, 1143

\bibitem[\protect\citeauthoryear{{Hoeft} \& {Br{\"u}ggen}}{{Hoeft} \&
  {Br{\"u}ggen}}{2007}]{HoeftBrueggen2007}
{Hoeft} M.,  {Br{\"u}ggen} M.,  2007, \mnras, 375, 77

\bibitem[\protect\citeauthoryear{{Kang}, {Ryu}, {Cen} \& {Ostriker}}{{Kang}
  et~al.}{2007}]{Kang2007}
{Kang} H.,  {Ryu} D.,  {Cen} R.,    {Ostriker} J.~P.,  2007, \apj, 669, 729

\bibitem[\protect\citeauthoryear{{Kang}, {Ryu} \& {Jones}}{{Kang}
  et~al.}{2012}]{Kang2012}
{Kang} H.,  {Ryu} D.,    {Jones} T.~W.,  2012, \apj, 756, 97

\bibitem[\protect\citeauthoryear{{Kauffmann} \& {White}}{{Kauffmann} \&
  {White}}{1993}]{Kauffmann1993}
{Kauffmann} G.,  {White} S.~D.~M.,  1993, \mnras, 261, 921

\bibitem[\protect\citeauthoryear{{Kirk} \& {Heavens}}{{Kirk} \&
  {Heavens}}{1989}]{Kirk1989}
{Kirk} J.~G.,  {Heavens} A.~F.,  1989, \mnras, 239, 995

\bibitem[\protect\citeauthoryear{{Kocevski}, {Ebeling}, {Mullis} \&
  {Tully}}{{Kocevski} et~al.}{2007}]{Kocevski2007}
{Kocevski} D.~D.,  {Ebeling} H.,  {Mullis} C.~R.,    {Tully} R.~B.,  2007,
  \apj, 662, 224

\bibitem[\protect\citeauthoryear{{Kronberg}}{{Kronberg}}{1994}]{Kronberg1994}
{Kronberg} P.~P.,  1994, Reports on Progress in Physics, 57, 325

\bibitem[\protect\citeauthoryear{{Lacey} \& {Cole}}{{Lacey} \&
  {Cole}}{1993}]{Lacey1993}
{Lacey} C.,  {Cole} S.,  1993, \mnras, 262, 627

\bibitem[\protect\citeauthoryear{{Macario}, {Markevitch}, {Giacintucci},
  {Brunetti}, {Venturi} \& {Murray}}{{Macario} et~al.}{2011}]{Macario2011}
{Macario} G.,  {Markevitch} M.,  {Giacintucci} S.,  {Brunetti} G.,  {Venturi}
  T.,    {Murray} S.~S.,  2011, \apj, 728, 82

\bibitem[\protect\citeauthoryear{{Markevitch}, {Govoni}, {Brunetti} \&
  {Jerius}}{{Markevitch} et~al.}{2005}]{Markevitch2005}
{Markevitch} M.,  {Govoni} F.,  {Brunetti} G.,    {Jerius} D.,  2005, \apj,
  627, 733

\bibitem[\protect\citeauthoryear{{Navarro}, {Frenk} \& {White}}{{Navarro}
  et~al.}{1996}]{nfw1996}
{Navarro} J.~F.,  {Frenk} C.~S.,    {White} S.~D.~M.,  1996, \apj, 462, 563

\bibitem[\protect\citeauthoryear{{Ogrean}, {Br{\"u}ggen}, {Simionescu},
  {R{\"o}ttgering}, {van Weeren}, {Croston} \& {Hoeft}}{{Ogrean}
  et~al.}{2013}]{Ogrean2013a}
{Ogrean} G.,  {Br{\"u}ggen} M.,  {Simionescu} A.,  {R{\"o}ttgering} H.,  {van
  Weeren} R.~J.,  {Croston} J.~H.,    {Hoeft} M.,  2013, Astronomische
  Nachrichten, 334, 342

\bibitem[\protect\citeauthoryear{{Ogrean}, {Br{\"u}ggen}, {R{\"o}ttgering},
  {Simionescu}, {Croston}, {van Weeren} \& {Hoeft}}{{Ogrean}
  et~al.}{2013}]{Ogrean2013b}
{Ogrean} G.~A.,  {Br{\"u}ggen} M.,  {R{\"o}ttgering} H.,  {Simionescu} A.,
  {Croston} J.~H.,  {van Weeren} R.,    {Hoeft} M.,  2013, \mnras, 429, 2617

\bibitem[\protect\citeauthoryear{{Ogrean}, {Br{\"u}ggen}, {van Weeren},
  {R{\"o}ttgering}, {Croston} \& {Hoeft}}{{Ogrean} et~al.}{2013}]{Ogrean2013d}
{Ogrean} G.~A.,  {Br{\"u}ggen} M.,  {van Weeren} R.~J.,  {R{\"o}ttgering} H.,
  {Croston} J.~H.,    {Hoeft} M.,  2013, \mnras, 433, 812

\bibitem[\protect\citeauthoryear{{Rossetti}, {Eckert}, {De Grandi},
  {Gastaldello}, {Ghizzardi}, {Roediger} \& {Molendi}}{{Rossetti}
  et~al.}{2013}]{Rossetti2013}
{Rossetti} M.,  {Eckert} D.,  {De Grandi} S.,  {Gastaldello} F.,  {Ghizzardi}
  S.,  {Roediger} E.,    {Molendi} S.,  2013, \aap, 556, A44

\bibitem[\protect\citeauthoryear{{Russell}, {Sanders}, {Fabian}, {Baum},
  {Donahue}, {Edge}, {McNamara} \& {O'Dea}}{{Russell}
  et~al.}{2010}]{Russell2010}
{Russell} H.~R.,  {Sanders} J.~S.,  {Fabian} A.~C.,  {Baum} S.~A.,  {Donahue}
  M.,  {Edge} A.~C.,  {McNamara} B.~R.,    {O'Dea} C.~P.,  2010, \mnras, 406,
  1721

\bibitem[\protect\citeauthoryear{{Skillman}, {Xu}, {Hallman}, {O'Shea},
  {Burns}, {Li}, {Collins} \& {Norman}}{{Skillman} et~al.}{2013}]{Skillman2013}
{Skillman} S.~W.,  {Xu} H.,  {Hallman} E.~J.,  {O'Shea} B.~W.,  {Burns} J.~O.,
  {Li} H.,  {Collins} D.~C.,    {Norman} M.~L.,  2013, \apj, 765, 21

\bibitem[\protect\citeauthoryear{{van Weeren}, {R{\"o}ttgering}, {Br{\"u}ggen}
  \& {Hoeft}}{{van Weeren} et~al.}{2010}]{vanWeeren2010}
{van Weeren} R.~J.,  {R{\"o}ttgering} H.~J.~A.,  {Br{\"u}ggen} M.,    {Hoeft}
  M.,  2010, Science, 330, 347

\bibitem[\protect\citeauthoryear{{Vazza}, {Br{\"u}ggen}, {van Weeren},
  {Bonafede}, {Dolag} \& {Brunetti}}{{Vazza} et~al.}{2012}]{Vazza2012}
{Vazza} F.,  {Br{\"u}ggen} M.,  {van Weeren} R.,  {Bonafede} A.,  {Dolag} K.,
   {Brunetti} G.,  2012, \mnras, 421, 1868

\bibitem[\protect\citeauthoryear{{Vazza}, {Brunetti}, {Kritsuk}, {Wagner},
  {Gheller} \& {Norman}}{{Vazza} et~al.}{2009}]{Vazza2009}
{Vazza} F.,  {Brunetti} G.,  {Kritsuk} A.,  {Wagner} R.,  {Gheller} C.,
  {Norman} M.,  2009, \aap, 504, 33

\end{thebibliography}

\appendix

\label{lastpage}

\end{document}